\documentstyle[preprint,12pt,aps]{revtex}     

\tighten
\begin{document}
\draft
\title{
\begin{flushright}
\begin{minipage}{3 cm}
\small
ZTF/99--12\\
\end{minipage}
\end{flushright}
LEPTON FLAVOR VIOLATION IN THE STANDARD MODEL EXTENDED BY HEAVY
SINGLET DIRAC NEUTRINOS}
\author{A.~Ilakovac \footnote{e-mail: ailakov@phy.hr}\\[.5cm]}
\address{University of Zagreb, Faculty of Science, Department of
Physics,  P.O.B 162,\\
Bijeni\v cka 32, 10001 Zagreb, Croatia\\[.3cm]}
\date{\today}

\maketitle

\begin{abstract}
Low energy neutrinoless 
lepton flavor violating (LFV) processes are studied
in an extension of the Standard Model (SM) by heavy $SU(2)\times U(1)$
singlet Dirac neutrinos.
An upper bound procedure 
is elaborated for the evaluation of amplitudes.
A comment on the extraction of heavy neutrino mixings from astrophysical 
observations is given. For processes not treated in the applied model 
the formalism for evaluating the branching ratios
(BRs) is presented.
The processes previously studied in the model are carefully examined.
Some of the previous results are improved. 
Special attention is paid to the structure of the amplitudes and
BRs as well as to the
relations between BRs of different LFV processes.
Numerical analysis of the BRs is done. 
The decoupling of heavy neutrinos is discussed and it 
is explicitely shown that the very heavy neutrinos decouple 
when the upper bound procedure is applied. 
The upper limits of
the BRs are compared with the current experimental upper
bounds and the processes interesting for the search 
for LFV are proposed.
The LFV decays are shown to be unsuitable for finding upper bounds on
"diagonal" LFV parameters.
The $B$-meson LFV processes 
are suggested for the search of LFV in future
$B$-factories. 
\end{abstract}


\section{Introduction}
\label{secI}

If the instanton effects \cite{tHooft76} are neglected, lepton flavor and
lepton number are conserved in the Standard Model (SM). 
Recently found atmospheric
neutrino oscillations \cite{SupKam98} indicate that neutrino masses 
are nondegenerate and the lepton flavor is not conserved. Independent 
confirmation of the deviation from SM is expected to manifest as
a nonconservation of lepton flavor/number (LFV/LNV), 
as a breaking
of lepton universality, in CP violating processes which are not
consistent with SM etc.

The problem of LFV/LNV is related   
to the physics beyond SM and
includes various physical areas 
\cite{DL95} :
atomic physics (e.g. muonium--antimuonium conversion),
nuclear physics ($\mu\to e$ conversion, double beta decay),
low energy hadron physics (leptonic and semileptonic decays
of mesons and tau lepton), problem of CP violation etc.

LFV have been found in various extensions of SM 
\cite{DL95,KLV94,LFV_SMext,V89}.
Here, LFV is studied 
within one of two extensions of SM by heavy neutrinos
\cite{AP_ZPC92,MV86}, obtained by 
adding additional heavy Dirac neutrinos to it. 
It is referred here as the $V$ model \cite{MV86}. Due to the 
Dirac character of the heavy neutrinos, there are no LNV processes in
this model.
The other model 
\cite{AP_ZPC92}, 
obtained by extending SM with additional heavy 
Majorana neutrinos, has some
renormalization problems and 
light neutrino mass problems \cite{KP96}. Besides the additional
heavy Dirac neutrinos, the $V$ model contains three massless neutrinos.
It should be noted that in this work the
$V$ model is used phenomenologically. Any model with the same
gauge properties and about equally large heavy neutrino masses would give
the same results, regardless whether the light neutrinos are massless or have 
masses which are in accord with the present experimental data.

The extensions of SM by heavy neutrinos contain a
Cabibbo-Kobayashi-Maskawa type matrix for leptons (LCKM). In general,
the elements of this matrix are not known. Experimental and theoretical
constraints exist only for some specific sums of the matrix elements
of the heavy 
neutrino part of the matrix. 
Therefore, 
the LFV amplitudes cannot be evaluated
exactly, but only the upper bounds on their values may be found
\cite{FI96}. 
The evaluation is especially complicated when the amplitudes contain
expressions with more than two LCKM matrix elements. 
In this paper a method 
for evaluation of the upper bounds of amplitudes found in the previous
publication \cite{FI96} is improved. The method
gives upper bounds for all values of the model parameters,
but in some directions
of the parameter space it is not very restrictive. 
It is explicitly shown that the upper bound procedure 
leads to the decoupling of
the heavy neutrinos in the infinite mass limit, showing that the
"nondecoupling" of heavy neutrinos \cite{IP_NPB,TBBJ95}
is only a transient effect, appearing with enlargement of the
heavy neutrino mass. 
It should be noted that this
"proof" of generalization of the Appelquist--Carrazone theorem is based
only on the requirement that the physical system can be described
pertubatively, and is independent of the introduction of somewhat
undetermined maximal 
$SU(2)_L$-doublet mass term as in Ref. \cite{TBBJ95}.  To give the feeling
how large error can be introduced using the upper bound procedure
elaborated here,
a few branching ratios (BRs)
obtained by the upper bound procedure are compared with BRs
obtained using "realistic" LCKM matrices.

The LFV processes are not very usefull for deriving
upper bounds on the matrix elements of LCKM matrix. The 
amplitudes for these processes are proportional to the 
sums of products of the LCKM matrix elements and functions of heavy neutrino
masses. Using the freedom 
to choose unknown phases of the LCKM matrix and heavy neutrino masses,
these sums can always be set to be equal zero,
even if the absolute values of nondiagonal elements of the LCKM matrix
are different from zero.
The present limits on the LCKM matrix elements are derived from the
measurements of lepton flavor conserving processes \cite{NRT94},
more precisely, from the estimates of deviations of the corresponding
decay rates from the SM results. For each row (a row corresponds to 
a specific lepton $l$) of the LCKM
matrix, these data 
give a limit on the sum of
squares of absolute values of the matrix elements corresponding to the
heavy neutrinos,
$(s_L^{\nu_l})^2$. Knowing the upper bounds on $(s_L^{\nu_l})^2$-s
one may derive the upper bound for BR of any LFV 
process. One of the aims of this paper is to derive the upper bounds 
of BRs for all low energy LFV processes in the $V$ model.
The processes having comparable theoretical and experimental 
upper bounds of the BR, 
or theoretical upper bound larger than the experimental one,
are interesting for further experimental investigation. 

Neutrino oscillations of two massless neutrinos 
in supernovae have been shown 
to give a very strong upper bound of two of the LCKM matrix elements 
in the part of the matrix corresponding to the massless neutrinos
\cite{NQRV96}. Here, the analysis has been repeated for three 
neutrinos, hoping that the upper bounds for 
other "massless neutrino" LCKM matrix elements
may be derived. The knowledge of nondiagonal "massless neutrino" LCKM
matrix elements may, in principle, lead to better upper bounds 
on some combinations of
"heavy neutrino" LCKM matrix elements than the terrestrial experiments. 
Unfortunately, the analysis made here shows that the three-neutrino
oscillations do not give  new constraints on any combination of "heavy
neutrino" LCKM matrix elements. The only new information it gives is
that the mixing between massless "mu" and "tau" neutrinos is smaller than 
the value obtained from the analysis of Super-Kamiokande data
\cite{SupKam98},
in which "mu" and "tau" neutrions were assumed to have small masses.

Untill now, many of the low energy neutrinoless 
LFV processes have been investigated.
Some of them were examined only within a few models, 
for instance LFV decays of
heavy mesons. The neutrinoless LFV
decays of B-mesons were studied in the frame of SM with additional Higgs
doublet \cite{SY91}, while the 
neutrinoless
LFV decays of D-mesons have been studied in the frame of leptoquark models
\cite{BW86} and a flipped left-right symmetric model \cite{JOE90}. Here,
they are analyzed in the $V$ model.
Some of the low energy LFV processes have not been studied in 
the frame of the $V$ model. Among them are
the muonium--antimuonium ($M\leftrightarrow\bar{M}$) 
conversion and  neutrinoless LFV violating 
decays of the $Z$ boson. The results are also given here. Some of 
the neutrinoless LFV processes have been analyzed in the $V$ model,
but the 
analysis is incomplete \cite{GGV92} or there are some 
errors in the expressions 
for amplitudes or decay rates \cite{FI96,IP_NPB,TBBJ95}. 
Here only the corrections 
to the previous results are given. 

On the quark and lepton level there are only a few Feynman diagrams 
(composite loop functions) that
contribute to any neutrinoless LFV decay amplitude. If two neutrinoless
LFV processes contain only one common composite loop function, the ratio
of corresponding BRs is independent of the $V$-model parameters.
Therefore, roughly speaking, knowing one
BR, BRs of processes comprising the same basic Feynman diagram
may be evaluated without the knowledge of parameters of the $V$ model.
If LFV decay amplitudes contain different loop functions or 
more loop functions, the ratio of the BRs depends on $V$-model
parameters.
Nevertheless, the mass dependence of the ratio of the BRs
simplifies in the limit of large heavy neutrino masses. Most of the
amplitudes become dependent essentially only on one of 
the composite loop functions.
In that limit, the ratios of the BRs having the same dominant 
composite loop
function become independent of the $V$-model parameters.
Experimentally,
for most neutrinoless LFV processes,
only the large heavy neutrino mass limit is interesting, because,
with few exceptions, only in that limit BRs
assume the values comparable with the
present day experimental limits. A comparitive analysis of the amplitudes 
and BRs of 
all neutrinoless LFV processes is presented.

In the section \ref{secCM}
some properties of the $V$ model, relevant for further discussion, are
given. Discussion on the limits of the model parameters is given 
in the section \ref{secLMPMEA}. 
The amplitudes for the neutrinoless LFV processes not
studied in the $V$ model, and some improvements and corrections of the
previous results are presented in section \ref{secNRLENLFVP}.
The amplitudes and BRs of LFV processes are studied in section
\ref{secLENLFVADR}.
The numerical results and comparison with experimantal limits
are also given in section \ref{secLENLFVADR}. 
Conclusions are summarized in section \ref{secC}.


\section{Comments on the Model} 
\label{secCM}
Here, a model with additional $SU_L(2)\times U(1)$ singlet Dirac neutrinos, 
which have large mixings with the SM leptons, is used in the calculations.
The masses of the singlet neutrinos are not restricted by the $SU_L(2)\times
U(1)$ breaking scale. The large mixings and the large masses are necessary
conditions for obtaining observable LFV decay rates.

In the model considered here \cite{MV86,BSVMV87,GGV89,WW83,Wi86},
the total lepton number
($L$) is conserved. For each SM neutrino 
one left-handed and one right-handed singlet neutrino is added, 
although, in
principle, the structure of the mass matrix permits addition of an
arbitrary number of pairs ($n_R$) of left-handed and right-handed
neutrinos ($Vn_R$ models).
Lepton number conservation gives such a structure to the
mass matrix which automatically leads to three massless neutrinos
at any order of the perturbation theory \cite{BSVMV87}.

Since the new neutrinos 
are $SU(2)_L\times U(1)$ singlets,
the structure of the lepton interaction vertices in the weak basis
is the same as in SM \cite{BSVMV87}. 
However, in a transition to the mass
basis, nondegeneracy of the
neutrinos leads to the Cabibbo-Kobayashi-Maskawa
(CKM) type matrix ($B_{ln}$)
in the charged current (CC) $nlW$  vertices. As only a part of the
mass-basis neutrinos interact with the $Z$ boson, neutral current (NC)
$nnZ$
vertices ($n$ is neutrino field in the mass basis) are also not
flavor-diagonal,
and contain matrix elements of the 
nondiagonal matrix ($C_{nn}$). The NC $llZ$ vertices and the quark
vertices are the same as in SM.

The $C$ matrix from the neutrino NC vertex may be expressed in terms
of
$B$ matrices from the CC lepton vertex. Therefore, besides the SM
parameters,
the model depends only on the 
$B$ matrix (or more precisely on the parameters defining the $B$ matrix)
and on heavy neutrino masses. The matrices $B$ and $C$ satisfy a set of
relations stemming from the gauge structure (see e.g. \cite{IP_NPB}),
\begin{eqnarray}
\label{BC}
\sum_{k=1}^{n_G+n_R}B_{l_1k}B^*_{l_2k}& =& \delta_{l_1l_2},\qquad
\sum_{k=1}^{n_G+n_R}C_{ik}C^*_{jk}\ =\ C_{ij},
\nonumber \\
\sum_{k=1}^{n_G+n_R}B_{lk}C_{ki}& =& B_{li},\qquad
\sum_{l=1}^{n_G}B^*_{li}B_{lj}\ =\ C_{ij}.\qquad
\end{eqnarray}
From the orthogonality relations for $B_{ln}$ 
matrix elements, phase arbitrariness
of leptons and $SU(n_R)$ invariance of massless neutrinos
lead to $n_Gn_R$ independent angles and 
$(n_G-1)(n_R-1)$ independent phases 
of the $B$ matrix \cite{RV90,BRV89}. Experimentally, only $n_G$
parameters $s_L^{\nu_l}$ may be estimated. Therefore, 
the $B$ matrix elements are undetermined even for the simplest
case with two additional heavy neutrinos, $n_R=2$.
Since the $B$ matrix elements are unknown,
the amplitudes of LFV processes cannot be evaluated 
exactly, but only upper bounds of the amplitudes may be found. 
One should mention that there exists a model
for which amplitudes of LFV processes can be evaluated
exactly, in the case of $n_R=2$ \cite{AP_ZPC92}. Unfortunately, 
as mentioned before, it is excluded because of some renormalization and 
light-neutrino mass problems.

The degeneracy of massless neutrinos
allows one to write the $B$ matrix in the following form 
\cite{V89,BSVMV87,DSGV90}
\begin{equation}
\label{B_mat}
B_{ln_k}=[(UD_A)_{l\nu_i},(UG)_{lN_I}],\qquad k=(i,I),
\end{equation}
where $U$ is a unitary matrix, $D_A$ is a diagonal matrix and G is a
matrix
satisfying $D_A^2+GG^\dagger=1$. Indices $i$ and $I$ denote massless and
massive neutrinos, respectively. From the structure of the $B$ matrix,
it follows that the massless neutrino CC in principle is not diagonal, 
leading to LFV \cite{BSVMV87,DSGV90,LL88_2}
and nonortogonal effective weak-neutrino states \cite{LL88_2}, although
neutrinos are massless.
On the other side, the massless-neutrino NC, which contains the $C$ matrix
elements, is diagonal \cite{BSVMV87}. 
Since there are no tree-level
flavor violating neutral currents (FCNCs) in the massless neutrino sector, the
universality of massless neutrino couplings is not satisfied, because,
in general, the elements of the diagonal matrix $D_A$ are not equal.
The nonuniversality of these couplings may have some astrophysical 
implications.

As mentioned in Introduction, the B matrices 
are used to define the parameters $s_L^{\nu_l}$, which are a measure
of the deviation from SM, in the following way 
\cite{TBBJ95,LL88,NRT92,BDGR91,BGKLM94}
\begin{equation}
\label{sL_def}
(s_L^{\nu_l})^2\ =\ \sum_{i=1}^{n_R}B_{lN_i}B^*_{lN_i}.
\end{equation}
Because the definition of $(s_L^{\nu_l})^2$ 
contains $B_{lN}$ matrix elements
of the same lepton flavor, the term 
"diagonal" mixing(s) will be sometimes used
in the text below.

\section{Limits on the model parameters and methods of evaluation of
amplitudes}
\label{secLMPMEA}
\subsection{Experimental limits}
The parameters $(s_L^{\nu_l})^2$ have been determined from the global
analysis of the low energy tree level processes 
\cite{NRT94,LL88,NRT92,BDGR91,BGKLM94}. In these processes
heavy neutrinos may manifest only indirectly, through change of the 
light (massless)-neutrino couplings. 
These couplings attain additional 
$c_L^{\nu_l}$ factors, where $(c_L^{\nu_l})^2\equiv 1-(s_L^{\nu_l})^2=
\sum_{i=1}^{n_G} B_{l\nu_i} B^*_{l\nu_i}$. The changes of the couplings
could show up as a nonuniversality of
CC couplings,
as a deviation from unitarity of the CKM matrix,
as a change of the invisible width of the $Z$
boson etc. \cite{NRT94,NRT92}. The best 
limits on the mixings $s_L^{\nu_l}$,
\begin{eqnarray}
\label{sL}
(s_L^{\nu_e})^2\ <\ 0.0071,\qquad 
(s_L^{\nu_\mu})^2\ <\ 0.0014,\qquad
(s_L^{\nu_\tau})^2\ <\ 0.033(0.01),
\end{eqnarray}
were found in Ref. \cite{NRT94}.
The value in the brackets is valid for $SU(2)_L\times U(1)$ singlet
heavy neutrinos.

The more stringent limits on the $B_{lN}$ matrix elements were searched
for investigating the loop effects in 
the lepton-conserving and lepton-violating processes. 
Direct limits on the parameters $s_L^{\nu_l}$ are not possible
as the expressions
derived from the loop
amplitudes, which are constrained by experimental data, depend not only on
the $s_L^{\nu_l}$ parameters but also on the $B_{lN}$ phases
and masses of heavy neutrinos.
Lepton conserving processes 
including heavy neutrinos in loops were studied by Kalyniak and Melo
\cite{BKM95,KM97}. They studied the loop effects of heavy Dirac
neutrinos on muon decay, universality-breaking ratio in $Z\to l\bar{l}$
decays and $\Delta r$ quantity. 
They found no new 
constraints on the $s_L^{\nu_l}$ parameters. 
The flavor nondiagonal (LFV) processes without light neutrinos in the final
state were studied extensively 
both theoretically \cite{DL95,KLV94,IP_NPB,BSVMV87,GGV89,RV90,LFVtheo,muexp} 
and experimentally \cite{DL95,tauexp,muexp,PDG98,KPJ98}. 
The advantage of these processes is that their
observation would be a clear and unambiguous signal for LFV.
These processes proceed only through loops. 
Using independence of the loop functions on the light neutrino masses
and the orthogonality
of rows of $B$ matrix, the amplitudes of these processes may
always be expressed
in terms of heavy neutrino contributions only.
Three of these processes, $\mu\to e\gamma$, $\mu\to 3e$ and $e$-$\mu$
conversion in $Ti$, gave new very stringent constraints on a specific
combinations of heavy neutrino masses and matrix elements 
$B_{eN}$ and $B_{\mu N}$ \cite{TBBJ95}.
Particularly, near-independence of the $\mu\to e\gamma$ amplitude on 
heavy neutrino masses enables
one to find the following very stringent mass independent limit,
\begin{equation}
\label{BeBmu}
\sum_{i=1}^{n_R} B^*_{\mu N_i}B_{eN_i}\ <\ 2.4\times 10^{-4}.
\end{equation}
No other constraints independent of heavy neutrino masses 
were derived from the LFV processes. It should be noted that the 
limit (\ref{BeBmu}) does not necessarily lead to new limits on
the $s_L^{\nu_l}$ parameters. The sum in (\ref{BeBmu}) may be written 
in terms of the parameters $s_L^{\nu_\mu}$ and $s_L^{\nu_e}$ and a complex
"cosine" of the "angle" between vectors $\{B_{\mu N_i}\}$ and $\{B_{e N_i}\}$,
\begin{equation} 
\label{xll0}
\sum_{i=1}^{n_R} B^*_{\mu N_i}B_{eN_i}\ =\ 
s_L^{\nu_\mu}s_L^{\nu_e}x^0_{\mu e},
\end{equation}
where $x^0_{\mu e}=\sum_{i=1}^{n_R} B^*_{\mu N_i}B_{eN_i}/
s_L^{\nu_\mu}s_L^{\nu_e}$. Obviously a reduction of $x^0_{\mu e}$
may assure the fulfillment of the inequality (\ref{BeBmu}) without reducing
the $s_L^{\nu_l}$ parameters.
Within the $V$ model the explicit 
estimates of BRs for the processes 
including more than two $B_{lN}$ matrices were given for the first time in
Ref. \cite{FI96}. 

\subsection{A comment on astrophysical limits}
The masslessness of "light" neutrinos in the $V$ model lead to 
the limits on some $B_{l\nu}$ matrix elements which can be
derived from astrophysical observations. Valle and collaborators
have noticed that the measurements of neutrino flux from the supernova
SN87 leads to two very small lepton--massless-neutrino mixings
\cite{NQRV96},
\begin{equation}
\label{astrolim}
|B_{e\nu_\tau}|,\ |B_{\tau\nu_e}|\  < 10^{-3}.
\end{equation}
The result (\ref{astrolim}) follows from an estimate of the 
$\nu_e$--$\nu_\tau$ conversion probability in the $V$ model.
To find whether similar upper bounds can be found for other
massless-neutrino
$B$ matrix elements, their calculation is repeated here for three
massless neutrinos. The motivation for such calculation is following.
Through the orthogonality relations for $B$ matrix elements (\ref{BC}),
very stringent limits on 
the matrix elements $B_{l\nu}$ would lead to better upper bounds on 
nondiagonal mixings $\sum_{i=1}^{n_R} B_{lN_i}B^*_{l'N_i}$ 
than those obtained by 
terrestrial experiments.

The derivation of the 
limits (\ref{astrolim}) is based on an analysis of neutrino
oscillations of the two neutrinos for which the experimental upper bounds 
on $s_L^{\nu_l}$ parameters are the weakest -- 
$s_L^{\nu_e}$ and $s_L^{\nu_\tau}$.
The oscillations of massless neutrinos are a consequence
of an interplay between the 
CC and NC neutrino weak interactions \cite{V87}. They 
appear only if the universality of the 
NC interactions is not fullfilled and if
the nondiagonal CC currents are different from zero. Following the notation
of Refs. \cite{NQRV96,V87}, the deviation from 
the universality is described by
small parameters $h_l$ (for small $h_l$, $h_l\approx s_L^{\nu_l}$). 
The massless-neutrino part of the $B$ matrix is parametrized by one 
mixing angle $\theta$, which is assumed to be small.
The resonance 
condition reads
\begin{equation}
2Y_e\ =\ \frac{h_\tau^2-h_e^2}{1+h_e^2},
\end{equation}
where $Y_e=n_e/(n_e+n_n)$, $Y_n=1-Y_e$ and $n_e$ and $n_n$ are 
the electron and the neutron number densities. As the experimental limits
on $h_e$ and $h_\tau$ are much smaller than one, the resonance condition 
can be fullfilled only in a highly neutronized medium, which can be found 
in supernovae explosions. In Ref. \cite{NQRV96} it was shown that the
neutrinosphere appears for the electron fraction $Y_e\approx 6\times10^{-3}$. 
The experimental upper bounds (\ref{sL}) show that the resonance condition
can be fullfilled for $Y_e\stackrel{\displaystyle <}{\sim} 0.015$, quite
close to the $Y_e$ value at the neutrinosphere. 
Assuming there is no nonforward scattering
of neutrinos \cite{Si95}, the authors of Ref. \cite{NQRV96} found the
probability for $\nu_e\leftrightarrow \nu_\tau$ and 
$\bar{\nu}_e\leftrightarrow \bar{\nu}_\tau$ conversions in a
simple Landau-Zener approximation \cite{LZapp,KP89},
\begin{eqnarray}
\label{Petau}     
P&\equiv&P(\bar{\nu}_e\to\bar{\nu}_\tau)\ =\ 
1-P(\bar{\nu}_e\to\bar{\nu}_e)\ =\ 
\frac{1}{2}-\bigg[\frac{1}{2}-
exp\bigg(-\frac{\pi^2}{2}\frac{\delta r}{L^{res}_m}\bigg)\bigg]
\cos 2\theta\cos 2\theta_m
\nonumber\\
& \approx& 1-exp\bigg(-\frac{\pi^2}{2}\frac{\delta r}{L^{res}_m}\bigg) 
\ \equiv 1-P_{LZ},
\end{eqnarray} 
where $P_{LZ}$ is the Landau-Zener crossing probability, 
$L_m^{res}$ is the neutrino oscillation length in matter at resonance, 
$\theta_m\approx\pi/2$ is the mixing angle in matter at production point
(neutrinosphere) and $\delta r=2\sin 2\theta |d\ln Y_e/dr|^{-1}_{res}$.
The approximate equality in (\ref{Petau}) is a conseqence of the small
mixing angle ($\theta$) approximation.
Using that result, the expression for the detected terrestrial flux 
\cite{SSB94},
\begin{equation}
\label{flux}
\phi_{\bar{\nu}_e}\ =\ \phi^0_{\bar{\nu}_e}(1-P)
 +\phi^0_{\bar{\nu}_\tau}P
\end{equation}
($\phi^0_{\bar{\nu}_e}$ and $\phi^0_{\bar{\nu}_\tau}$ are 
$\bar{\nu}_e$ and $\bar{\nu}_\tau$ fluxes in the absence of the neutrino
conversion, respectively), the model independent result for the
probability for $\bar{\nu}_e\leftrightarrow \bar{\nu}_\tau$ conversion, 
$P<0.35$ \cite{SSB94},
and the density profiles for $Y_e$ from the Wilson supernova 
model, Valle and his collaborators
found the result given in Eq. (\ref{astrolim}).

Following the procedure of Ref. \cite{NQRV96}, a similar analysis can be
done for the three massless neutrinos. To analyze the terrestrial flux 
data, one should know only the survival probability of the electron
antineutrino $P(\bar{\nu}_e\to\bar{\nu}_e)$ \cite{KP89,KP88}. 
The Eq. (\ref{flux}) is still valid, but $\phi^0_{\bar{\nu}_\tau}$
represents the sum of $\bar{\nu}_\mu$ and $\bar{\nu}_\tau$ fluxes.
In the three-neutrino case there are two
resonances: $\bar{\nu}_e\leftrightarrow \bar{\nu}_\mu$ resonance and
$\bar{\nu}_e\leftrightarrow \bar{\nu}_\tau$ resonance. According to 
the limits (\ref{sL}) and the $Y_e$ value at the neutrinosphere, the 
$\bar{\nu}_e\leftrightarrow \bar{\nu}_\mu$ resonance is within the 
neutrinosphere. Therefore, the effects of this resonance do not
contribute to 
$P(\bar{\nu}_e\to\bar{\nu}_e)$. 
Taking that into account (or equivalently taking the
neutrinosphere as a source of neutrinos) and using the approximative
Kuo--Pantaleone treatment for three neutrino 
oscillations \cite{KP89} adjusted for physical situation studied here,
one obtains the following expression for $P\equiv
1-P(\bar{\nu}_e\to\bar{\nu}_e)$,
\begin{eqnarray}
P& =& 1-(|U_{e1}|^2P_{LZ}+(1-P_{LZ})|U_{e2}|^2))(|U_{e1}|^2+|U_{e2}|^2)
-|U_{e3}|^4
\nonumber\\
& =& 1-P_{LZ}\, \cos^4\phi \cos 2\omega 
-\cos^4\phi\sin^2\omega-\sin^4\omega
\end{eqnarray}
(neutrino states 1, 2 and 3 are mainly $\nu_e$, $\nu_\tau$ and
$\nu_\mu$ flavor states, respectively; 
the angles $\omega$ and
$\phi$ perform rotations between $1$ and $2$ states, and $2$ and $3$ 
states, respectively).
The Landau-Zener crossing probability $P_{LZ}$ can be obtained from the 
$P_{LZ}$ for two-neutrino oscillations, replacing $\sin 2\theta$ with 
$2U_{e1}U_{e2}=\cos^2\phi\sin 2\omega$ in the two-neutrino $P_{LZ}$.
In the small angle approximation, assumed in Ref. \cite{NQRV96},
the probability $P$ tends to zero only if $P_{LZ}$ is almost equal
one. Using the result of Ref. \cite{SSB94} mentioned above, $P<0.35$,
the small angle approximation,
and the analysis of Ref. \cite{NQRV96}, one finds limits on mixing
angles $\omega$ and $\phi$
\begin{equation}
\sin^22\omega<1\times 10^{-6}, \qquad \phi^2<0.27.
\end{equation}
The first limit corresponds to the limit obtained in the two neutrino
case. The second one is too weak to give limits on the $B_{lN}$ matrix
elements. Therefore, astrophysical measurements give no new limits on
the heavy neutrino part of the $B$ matrix.

The second limit has to be compared with the $\nu_\mu$--$\nu_\tau$ 
mixing angle obtained from the favorite interpretation of 
recent Super-Kamiokande results \cite{SupKam98},
$\theta_{\nu_\mu\nu_\tau}\approx \pi/4$.
Obviously, these two results are in a slight contradiction.

\subsection{Theoretical limits}
\label{secThelim}
If one wants to work in the perturbative regime of the
theory, an additional constraint on the $B_{lN}$ mixings comes from the 
theoretical argument that the partial wave unitarity 
(perturbative unitarity) has to be 
satisfied. 
From the perturbative unitarity follows that the 
decay width of any heavy neutrino has to be smaller than a half of 
its mass. Written in terms of heavy neutrino masses and 
$B_{lN}$-s, this condition reads \cite{FI96}
\begin{equation}
\label{PUB}
m_{N_i}^2\sum_{j=1}^{n_G}|B_{l_jN_i}|^2\leq
\frac{4}{\alpha_W}M_W^2\ \equiv\ m_D^2.
\end{equation}
$m_D$ represents the upper value the Dirac mass may 
attain in the neutrino mass matrix.
The perturbative unitarity bound (PUB) inequalities (\ref{PUB}) 
give upper limit
on a combination of a heavy neutrino mass $m_{N_i}$ and the matrix
elements $B_{lN_i}$.
Using Eq. (\ref{sL_def}), these relations may be combined into the limit
for the lightest heavy neutrino mass
\begin{equation}
\label{PUB_PV}
m_{N_1}^2\leq (m_{N_1}^0)^2(1+\sum_{i=2}^{n_R}\rho_i^{-2}),
\end{equation}
where
$(m_{N_1}^0)^2=4M_W^2/(\alpha_W\sum_{j=1}^{n_G}(s_L^{\nu_j})^2)$
and $\rho_i=m_{N_i}/m_{N_1}$.
Concerning the calculation of BRs,
the bound is very effective if the heavy neutrino
masses are equal. If the heavy neutrino masses differ considerably,
the bound is not very restrictive.
Namely, if one of the
heavy neutrino masses is smaller than $m_{N_1}^0/n_R^{1/2}$, 
the others may acquire infinite
values not followed by infinitely small values of the corresponding 
$B_{lN_i}$ mixings.
That leads to divergent BRs.
Therefore, one has to use the original inequality (\ref{PUB})
to restrict
model parameters. One cannot obtain closed
expressions since the model has too many free parameters, but one
can write two very rough bounds \cite{FI96}
\begin{eqnarray}
\label{PUB_APP}
|B_{lN_i}|&\leq &s_L^{\nu_i},
\nonumber\\
|B_{lN_i}|&\leq &\frac{2M_W}{\alpha_W^{1/2}m_{N_i}}
\ \equiv\ B_{lN_i}^{(0)},
\end{eqnarray}
originating from Eqs. (\ref{sL_def}) and (\ref{PUB}), respectively,
which have to be satisfied simultaneously. If the heavy neutrino
masses differ considerably, the bounds (\ref{PUB_APP}) are better
for finding upper bounds of BRs than the bound (\ref{PUB_PV}).

The "realistic" $B_{lN_i}$-s which 
automatically satisfy the PUB-s (\ref{PUB_APP}) and fullfill the
relation $\sum_i B_{lN_i}B_{lN_i}^*\leq (s_L^{\nu_l})^2$ may be obtained
by putting
\begin{equation}
\label{B_ex}
B_{lN_i}=((s_L^{\nu_l})^{-1}+(B_{lN_i}^{(0)})^{-1})^{-1}n_R^{-1/2}.
\end{equation}
This choice of $B_{lN_i}$-s is used below to give an
estimate how large error can be done in the evaluation of BRs
using the rough 
upper bound procedure presented above.
The $B_{lN_i}$ defined in 
Eq. (\ref{B_ex}) begins to differ considerably from the value
$s_L^{\nu_i}$ for
$m_{N_i}\stackrel{\displaystyle >}{\sim} 100M_W(0.1/s_L^{\nu_i})$.
Therefore, for $m_{N_i}$ values smaller than $2000\ GeV$, the 
$B_{lN_i}$-s are determined by experimental upper bounds (\ref{sL})
and not by the theoretical PUB limits $B_{lN_i}^{(0)}$. 

\subsection{Upper bound procedure for LFV amplitudes}
\label{secUBPFLFVA}
The equations (\ref{PUB_APP}) and (\ref{B_ex}) are the basis 
for evaluation of the LFV amplitudes. The evaluation 
based on Eq. (\ref{PUB_APP}) gives the upper bounds on 
absolute values of the amplitudes \cite{FI96}, 
which have to be satisfied by
any model with additional heavy neutrinos. It uses the Schwartz's inequality
for the product of two vectors. 
It always gives larger estimates
for an amplitude than the approach based on Eq. (\ref{B_ex}).
In both approaches the phases of the $B_{lN_i}$-s are
neglected, but in a different manner.
In the first approach the upper bound value of the amplitude is formed,
while in the second the $B_{lN_i}$-s are taken to be 
real and positive.
Both approaches explicitely show 
that the very heavy neutrinos are decoupled.
That is, they have no influence on the amplitudes of low energy 
LFV processes, in accord with the Appelquist-Carazzone theorem and its
generalization \cite{AC,SS}. 

Here, the
improved version of the upper bound procedure introduced in Ref. \cite{FI96}
is given.
The low energy LFV amplitudes may be written in terms of
\begin{equation}
\label{UL}
\sum_{i=1}^{n_R} B_{lN_i}^*B_{l'N_i} f(N_i,\cdots),\qquad
\sum_{j=1}^{n_G} V_{u_jd_a}V_{u_jd_a}^* f(u_j,\cdots)\quad
\mbox{and}\quad
\sum_{j=1}^{n_G} V_{ud_j}^*V_{ud_j} f(d_j,\cdots),
\end{equation}
where $f(N_i,\cdots)$, $f(u_j,\cdots)$ and $f(d_j,\cdots)$
are expressions comprising the loop functions.
The dots represent the indices not written
explicitly. Namely, the amplitudes often contain more 
than one sum over neutrino or quark flavors. Using the inequalities that
can be derived from Schwartz's inequality,
\begin{eqnarray}
\label{aibici1}
|\sum_i a_ib_ic_i|& \leq& \sum_i|a_i||b_i||c_i|,
\\
\label{aibici2}
|\sum_{i=1}^n a_ib_ic_i|& \leq& |{\bf a}||{\bf b}|\langle c\rangle+
|{\bf a}||{\bf b}|(\sum_{i=1}^n|c_i-\langle c\rangle|^2)^{1/2},
\end{eqnarray}
($\langle c\rangle = \sum_{i=1}^n c_i/n$)
and definition of $s_L^{\nu_l}$ (\ref{sL_def}), 
one can write the following upper limits for the expressions
(\ref{UL}),
\begin{eqnarray}
\label{SI1}
\Big|\sum_{i=1}^{n_R} B_{lN_i}^*B_{l'N_i} f(N_i,\cdots)\Big|& \leq&
s_L^{\nu_l}s_L^{\nu_l'}\Big(|\langle f(\cdots)\rangle_N|
+[\sum_{i=1}^{n_R}(f(N_i,\cdots)-\langle f(\cdots)\rangle_N)^2]^{1/2}
\Big),
\\
\label{SI2}
\Big|\sum_{j=1}^{n_R} V_{u_jd_a}V_{u_jd_a}^* f(u_j,\cdots)\Big|
& \leq& \sum_{j=1}^{n_R} |V_{u_jd_a}||V_{u_jd_a}||f(u_j,\cdots)|,
\nonumber\\
\Big|\sum_{j=1}^{n_R} V_{ud_j}V_{ud_j}^* f(d_j,\cdots)\Big|
& \leq& \sum_{j=1}^{n_R} |V_{ud_j}||V_{ud_j}||f(d_j,\cdots)|,
\end{eqnarray}
where $\langle\ \rangle_N$ represents the average over heavy neutrinos.
The inequality (\ref{aibici1}) gives the best estimate for the upper limit
if the components $c_i$ differ considerably. 
The inequality (\ref{aibici2}) gives the better estimate of the upper bound
if the components $c_i$ are approximately equal. As the amplitudes 
$f(u_j,\cdots)$ and $f(d_j,\cdots)$ depend strongly on quark masses, 
Eqs. (\ref{SI2}) give good estimates
for the upper bounds. Eq. (\ref{SI1}) is effective if the 
heavy neutrino masses are nearly degenerate, because most of the
$f(N_i,\cdots)$ functions depend strongly on the heavy neutrino masses. 
If one or more heavy neutrino masses differ considerably from the others,
then Eq. (\ref{SI1}) may lead even to a divergent result as the heavy
neutrino mass(es) tend to infinity. To avoid such undesirable behavior,
one has to use the combination of the upper bounds (\ref{aibici1})
and (\ref{aibici2}) for each set of heavy neutrino mass values
in the following manner. First, the heavy neutrino
masses are arranged in increasing order. The arranged masses
are devided into two sets, one containing smaller masses and the other
larger masses. There are $J+1$ such partitions, where $J$ is a number of
different heavy neutrino masses. Then $J+1$ different upper bounds of
the expression $\sum_{i=1}^{n_R} B_{lN_i}^*B_{l'N_i} f(N_i,\cdots)$ are
formed combining the upper bounds (\ref{aibici1}) and (\ref{aibici2}),
\begin{eqnarray}
\label{UB_part}
\Big|\sum_{i=1}^{n_R} B_{lN_i}^*B_{l'N_i} f(N_i,\cdots)\Big|& \leq&
s_L^{\nu_l}s_L^{\nu_l'}\Big(|\langle f(\cdots)\rangle_s|
+[\sum_{i_s}(f(N_{i_s},\cdots)-\langle f(\cdots)\rangle_s)^2]^{1/2}
\Big)
\nonumber\\
&&+\sum_{i_b}B^0_{lN_{i_b}}B^0_{l'N_{i_b}}|f(N_{i_b},\cdots)|,
\end{eqnarray}
where $\sum_{i_s}$ sums over the lighter heavy neutrino masses, and 
$\sum_{i_b}$ over the heavier ones. Finally, the numerical values of 
the $J+1$ upper bounds (\ref{UB_part}) are compared and the smallest of them
is taken to be the upper bound value. For amplitudes containing sums 
over two (heavy neutrino and/or quark) indices, the procedure is 
essentially the same. Again one looks for the minimal upper bound
value between upper bounds obtained for all possible partititions of heavy
neutrino masses.
This procedure gives convergent results for absolute 
values of the amplitudes, and it leads 
to the decoupling of the very heavy neutrinos.

It should be noted that the above upper bound procedure
gives {\bf upper bounds} for BRs
for neutrinoless LFV processes. 
Recently,  {\bf lower bound} limits for $\tau$ lepton decays were 
found using the Super-Kamiokande data on atmospheric
deficit of $\nu_\mu$, and interpreting it in terms of the best fit
to these data \cite{XYP98}. The mild GIM mechanism suppression, 
coming from logarithmic dependence on light neutrino masses,
appearing in $\tau\to \mu l^+l^-/\mu\rho^0$ decays,
leads to the lower bounds of the 
BRs
as large as $\sim 10^{-14}$. As the experimental upper limits on these
processes are of the order of $\sim 10^{-6}$ this lower limit  
is very welcome, because it strongly restricts the window
for the heavy neutrino LFV effects. However, these results have to 
be taken with caution, as the standard interpretation of the
Super-Kamiokande data is not the only one \cite{GGetal98}, although 
recent papers \cite{FLMS99,FGGV99} 
showed that the energy dependence of the oscillation
wavelength strongly supports the standard interpretation. 
It should be noted that the used $V$ model can 
easily be modified to include masses for massless neutrinos 
\cite{GGV89}. The results
for the neutrinoless LFV decays 
almost do not change if light neutrino masses,
consistent with Super-Kamiokande measurements, are introduced.

\section{New results on low-energy neutrinoless LFV processes}
\label{secNRLENLFVP}
As mentioned in Introduction, heavy meson neutrinoless LFV decays
and $M\leftrightarrow\bar{M}$ conversion have
not been studied in the $V$ model. They are examined below.
Some previous results for neutrinoless LFV decays 
are extended and/or corrected.

\subsection{Neutrinoless LFV decays of heavy mesons}
\label{secNLFVDOHM}
The LFV decays of heavy mesons were discussed in a few papers
in the context of the leptoquark models \cite{BW86}, a flipped left-right
symmetric model \cite{JOE90} and SM with an additional Higgs doublet 
\cite{SY91}. 

In these decays both lepton and quark flavor are
changed. In the $V$ model they can proceed only through 
box diagrams in which two $W$ bosons are exchanged. 
The effective Lagrangian on the quark-lepton level reads
\begin{eqnarray}
\label{Leff}
{\cal L}_{eff}& =& \frac{\alpha_W^2}{16M_W^2}\,
\sum_{l\neq l'}\sum_Q\sum_{q_a}F_{box}^{l'lq_aQ}\;
\bar{l}\gamma_\mu(1-\gamma_5)l'
\nonumber\\&&
\bar{q}_a\gamma^\mu(1-\gamma_5)Q\;
[\delta_{Qc}\delta_{q_au}-\delta_{Qb}(\delta_{q_ad}+\delta_{q_as})].
\end{eqnarray}
$l$ and $l'$ are the lepton fields, 
$q_a$ and $Q$ are the light and heavy quark
fields, respectively, $\alpha_W$ is 
the weak fine-structure constant, $M_W$ is
the $W$ boson mass and $F_{box}^{l'lq_aQ}$ is the composite loop function,
\begin{eqnarray}
\label{FHbox}
F_{box}^{l'luc}& =& \sum_{i=1}^{n_R}\sum_{j=1}^{n_G}B^*_{l'N_i}B_{lN_i}
V^*_{ud_j}V_{cd_j}\;
[H_{box}(\lambda_{N_i},\lambda_{d_j})
\nonumber\\&& 
-H_{box}(\lambda_{N_i},0)
-H_{box}(0,\lambda_{d_j})
+H_{box}(0,0)],
\nonumber\\
F_{box}^{l'lq_ab}& =& \sum_{i=1}^{n_R}\sum_{j=1}^{n_G}B^*_{l'N_i}B_{lN_i}
V_{u_jd_a}V^*_{u_jb}\;
[F_{box}(\lambda_{N_i},\lambda_{u_j})
\nonumber\\&&
-F_{box}(\lambda_{N_i},0)
-F_{box}(0,\lambda_{u_j})
+F_{box}(0,0)].
\end{eqnarray} 
$F_{box}$ and $H_{box}$ are loop functions defined in \cite{IKP95}.
These loop functions have approximatively 
logarithmic dependence on the heavy neutrino masses.

The dominant processes are those which have 
maximal value of the CKM matrix elements, 
maximal LCKM matrix elements and $t$-quark
mass in the loop function. The main neutrinoless LFV candidates,
between the two-prong and three-prong
processes studied here, are
$\bar{B}_s^0\to \tau^\pm e^\mp$, 
$B^-\to K^-\tau^\pm e^\mp$, 
$\bar{B}^0\to \bar{K}^0\tau^\pm e^\mp$ and 
$\bar{B}_s^0\to \phi\tau^\pm e^\mp$. There are no interesting
$D$ meson candidates for two reasons. One is of dynamical origin --
the quark masses 
involved in loop functions are smaller than in B-meson decays, so loop
functions are much smaller. The only larger loop contribution coming
from the b-quark is suppressed by small CKM matrix elements. The other is 
kinematical -- the difference of $\tau$ lepton and
$D$ meson masses is small. The small quark masses in loops and large 
t-quark width makes 
LFV decays of t-quark uninteresting from the experimental point of
view.

The matrix element of the neutrinoless 
LFV decay of a heavy meson $H$,
$H\to Xll'$, contains hadronic matrix element
$\langle X|\bar{q}_a(0)\gamma^\mu(1-\gamma_5)Q(0)|H\rangle$. The
corresponding matrix elements are usually parametrized in the
following way \cite{Cas97,BFO95}
\begin{eqnarray}
\langle 0|\bar{q}_a(0)\gamma_\mu(1-\gamma_5)Q(0)|H_a(p)\rangle& =& 
-if_H p_\mu ,
\nonumber\\
\langle P(p')|\bar{q}_a(0)\gamma_\mu(1-\gamma_5)Q(0)|H_a(p)\rangle& =&
\bigg[\bigg((p+p')_\mu-\frac{m_H^2-m_P^2}{q^2}q_\mu\bigg)F_1(q^2)
\nonumber\\&&
+\frac{m_H^2-m_P^2}{q^2}q_\mu F_0(q^2)\bigg]N_P^{q_a},
\nonumber\\
\langle V(p',\varepsilon)|
\bar{q}_a(0)\gamma^\mu(1-\gamma_5)Q(0)|H_a(p)\rangle& =&
\bigg[-\frac{2V(q^2)}{m_H+m_V}\varepsilon^{\mu\nu\alpha\beta}
\varepsilon^*_\nu p_\alpha p'_\beta
-i\varepsilon^*\cdot q\frac{2m_V}{q^2}q_\mu A_0(q^2)
\nonumber\\&&
-\frac{i\varepsilon^*\cdot q}{m_H+m_V}
\bigg((p+p')_\mu-\frac{m_H^2-m_V^2}{q^2}q_\mu\bigg)A_2(q^2)
\nonumber\\&&
+i(m_H+m_V)\bigg(\varepsilon^*_\mu
  -\frac{\varepsilon^*\cdot q}{q^2} q_\mu\bigg)A_1(q^2)\bigg]N_V^{q_a}.
\end{eqnarray}
$H_a$ is a heavy pseudoscalar meson containing light quark $\bar{q}_a$, 
$P$ and $V$ are a light pseudoscalar
meson and a light vector meson, respectively, $p$ and $p'$ 
are 4-momenta of the heavy and light meson, respectively, 
$q=p-p'$ is the momentum transfer, $\varepsilon$ is
the polarizaton vector of the light vector meson, 
$f_H$ is the decay constant of the heavy pseudoscalar meson, $F_1$, $F_2$, 
$V$, $A_0$, $A_1$ and $A_2$ are form factors and $N_P^{q_a}$ 
($N_V^{q_a}$) is a factor in front of the term containing $\bar{q}_a$ 
in the quark wave function of the $P$ ($V$) meson.
The $q^2$ dependence of the form factors is a consequence
of long-distance (resonance) effects following from strong interactions.

To evaluate the hadronic matrix elements of quark currents and to include
the long distance effects, 
one has to express the quark currents in
terms of the meson states and to introduce a strong-interaction Lagrangian
on the meson level. 
Similar hadronic matrix elements have been
extensively studied in radiative, semileptonic and nonleptonic decays
of heavy mesons. The combination of heavy quark effective theory (HQET)
and chiral pertubation theory (CHPT) has
been applied to these decays \cite{HQ_CH}. 
Here, the modification of this formalism
\cite{BFO95,BFO96a,BFO96b,BFOP97} is used. The 
authors of these papers replaced the HQET propagators by 
the full heavy-quark propagators, and introduced  $SU(3)$ 
symmetry breaking through physical masses and decay constants
of light mesons. The matrix elements in that approach read
\begin{eqnarray}
\label{hadr}
\langle 0|\bar{q}_a(0)\gamma_\mu(1-\gamma_5)Q(0)|{\cal H}(p)\rangle& =&
-if_{\cal H} p_\mu ,
\nonumber\\
\langle P(p')|\bar{q}_a(0)\gamma_\mu(1-\gamma_5)Q(0)|{\cal H}(p)\rangle& =&
N_P^{q_a}\bigg[-\frac{f_{\cal H}}{f_P}p_\mu
+2\frac{f_{{\cal H}'^*}}{f_P}(m_{\cal H}m_{{\cal H}'^*})^{1/2}
\nonumber\\&&
\times g\bigg(p'_\mu-\frac{p'\cdot q q_\mu}{m_{{\cal H}'^*}^2}\bigg)
\frac{m_{{\cal H}'^*}}{q^2-m_{{\cal H}'^*}^2}\bigg],
\nonumber\\
\langle V(p',\varepsilon)|
\bar{q}_a(0)\gamma^\mu(1-\gamma_5)Q(0)|{\cal H}(p)\rangle& =&
N_V^{q_a}\bigg[2^{3/2}\lambda g_V
  \bigg(\frac{m_{{\cal H}'^*}}{m_{\cal H}}\bigg)^{1/2}f_{H'^*}
  \frac{m_{{\cal H}'^*}}{q^2-m_{{\cal H}'^*}^2}\varepsilon_{\mu\nu\alpha\beta}
  \varepsilon^{*\mu}p^\alpha p'^\beta
\nonumber\\&&
-i\,2^{1/2}\beta g_V\bigg(\frac{m_{{\cal H}'}}{m_{\cal H}}\bigg)^
{1/2}f_{{\cal H}'}
  \frac{q\cdot\varepsilon^* q_\mu}{q^2-m_{\cal H}'^2}
-i\,2^{1/2}\alpha_1 g_V m_{{\cal H}'}^{1/2}\varepsilon^{*\mu}
\nonumber\\&&
+i2^{1/2}\alpha_2g_V m_{{\cal H}'}^{1/2}
  \frac{p_\mu p\cdot\varepsilon^*}{m_{\cal H}^2}\bigg].
\end{eqnarray}
${\cal H}'$ and ${\cal H}'^*$ 
represent heavy pseudoscalar meson and heavy vector
meson resonances, respectively, $f_{{\cal H}'}$, 
$f_{{\cal H}'^*}$, $m_{{\cal H}'}$, $m_{{\cal H}'^*}$ 
are the corresponding decay constants and masses, 
$g_V$~($\approx 6.0(2/a)^{1/2}$ with $a=2$ in the case of exact vector meson
dominance) is the vector meson self-interaction coupling constant
\cite{BKY88}, $g$ and $\beta$ are the coupling constants in the even 
part of the strong interaction 
Lagrangian \cite{Cas97,BFO95,BFO96a,BFOP97,Cas92,Dgamma}, 
$\lambda$ is a coupling
constant in the odd part of the strong 
interaction Lagrangian
\cite{Cas97,BFO95,BFO96a,BFO96b,BFOP97,Cas92}, and $\alpha_1$ and
$\alpha_2$ are coupling constants in the definition of weak current
\cite{BFO96a,BFOP97}.
The constants $g$, $\beta$ 
$\lambda$, $\alpha_1$ and $\alpha_2$
are free parameters which have to be determined from experimental data. 
 
The matrix elements of ${\cal H}\to Xll'$ follow from
(\ref{Leff}), (\ref{FHbox})
and (\ref{hadr}). From these matrix elements follow the corresponding 
decay rates:
\begin{eqnarray}
\label{BRHXll}
B({\cal H}_a^0\to l^-l'^+)& =& \frac{\alpha_W^4}{2^{10}\pi }
\frac{f_{{\cal H}^0}^2m_{{\cal H}^0}^3}
{\Gamma_{{\cal H}^0}M_W^4}\,
\frac{\lambda^{1/2}(m_{{\cal
H}^0}^2,m_{l'}^2,m_l^2)}{m_{{\cal H}^0}^2}\,
\frac{m_{{\cal H}^0}^2(m_{l'}^2+m_l^2)-(m_{l'}^2-m_l^2)}
{m_{{\cal H}^0}^4}\, |F_{box}^{l'lq_aQ}|^2,
\nonumber\\
B({\cal H}_a\to Pl^-l'^+)& =& 
\frac{\alpha_W^4(N_P^{q_a})^2}
{2^{13}\pi^3}\,
\frac{\int_{(m_l+m_{l'})^2}^{(m_{{\cal H}}-m_P)^2}dt\,
[a_P^2Z_{P1}+a_Pb_PZ_{P2}+b_P^2Z_{P3}]}
{m_{{\cal H}}^3\Gamma_{{\cal H}} M_W^4}\,
|F_{box}^{l'lq_aQ}|^2,
\nonumber\\
B({\cal H}_a\to Vl^-l'^+)& =& 
\frac{\alpha_W^4(N_V^{q_a})^2}
{2^{12}\pi^3}\,
|F_{box}^{l'lq_aQ}|^2
\frac{1}{m_{{\cal H}}^3\Gamma_{{\cal H}} M_W^4}
\int_{(m_l+m_{l'})^2}^{(m_{{\cal H}}-m_V)^2}dt\,
\Big[a_V^2Z_{V1}+b_V^2Z_{V2}+c_V^2Z_{V3}
\nonumber\\
&&+d_V^2Z_{V4}+a_Vc_VZ_{V5}+b_Vc_VZ_{V6}+b_Vd_VZ_{V7}
+c_Vd_VZ_{V8}\Big].
\end{eqnarray}
The form factors $a_P$, $b_P$, $a_V$, $b_V$ $c_V$ and $d_V$, and 
phase functions $Z_{Pi}$, $i=1,2,3$ and $Z_{Vi}$ $i=1,\cdots ,8$ are defined 
in Appendix.

\subsection{Muonium--antimuonium conversion}
The CC vertices in the $V$ model have $V-A$ structure. 
The effective Lagrangian for the $M\leftrightarrow\bar{M}$
conversion comes from the lepton box amplitude.
Therefore, the structure of the effective Hamiltonian density
for $M\leftrightarrow\bar{M}$ has the same $(V-A)\times(V-A)$ 
form as in the 
Feinberg's and Weinberg's papers \cite{FW61}
\begin{equation}
\label{HMM}
{\cal H}\ =\ G_{M\bar{M}}\;\bar{\psi}_\mu\gamma_\lambda(1-\gamma_5)\psi_e
\bar{\psi}_\mu\gamma^\lambda(1-\gamma_5)\psi_e,
\end{equation}
in which they had elaborated the original idea of Pontecorvo \cite{Po57}.
The constant $G_{M\bar{M}}$ contains information on physics beyond
SM. In the frame of the $V$ model it comprises 
the parameters of the box amplitude for the process 
$\mu^+e^-\to\mu^-e^+$,
which is forbidden in SM,
\begin{equation}
G_{M\bar{M}}\ =\ \frac{\alpha_W^2}{16 M_W^2}
F_{box}^{\mu ee\mu}.
\end{equation}
$F_{box}^{\mu ee\mu}$ is a composite loop function 
having the following structure \cite{IP_NPB}
\begin{equation}
\label{Fboxmeem}
F_{box}^{\mu ee\mu}\ =\ 2\sum_{ij=1}^{n_R}
B_{eN_i}B_{eN_j}B_{\mu N_i}^*B_{\mu N_j}^*
[F_{box}(\lambda_{N_i},\lambda_{N_j})-F_{box}(0,\lambda_{N_j})
-F_{box}(\lambda_{N_i},0)+F_{box}(0,0)].
\end{equation}
Using the expression (\ref{Fboxmeem}) 
for large degenerate heavy neutrino masses,
one obtains the limit 
\begin{equation}
\label{GMMth}
G_{M\bar{M}}\ \leq\ 3.9\times 10^{-5} x^0_{\mu ee\mu}\; G_F,
\end{equation}
where $G_F$ is the Fermi constant and $x^0_{\mu ee\mu}=F_{box}^{\mu ee\mu}/
(0.5\lambda_N(s_L^{\nu_\mu})^2(s_L^{\nu_e})^2)$. From the definition of 
the composite loop function and the limit (\ref{BeBmu}) follows that 
the $x^0_{\mu ee\mu}$ may assume only values smaller
than $4.7\times 10^{-3}$. Keeping that in mind, the 
result (\ref{GMMth}) has to be compared with 
the recent experimental upper 
bound \cite{KPJ98,LW99} which improved the previous 
experimental result \cite{M91G94} 
by the factor $\sim 50$, $G_{M\bar{M}}\leq 3.0\times
10^{-3}G_F$. The upper bound (\ref{GMMth}) is larger than the 
result found by Swartz \cite{MLS89}, estimated  within SM
with massive 
Dirac neutrinos, by comparing the effective Hamiltonians for
$M$-$\bar{M}$ conversion and for $B^0$-$\bar{B}^0$ transition. 
Having in mind that the upper limit (\ref{BeBmu}) was
much weaker than when Swartz wrote his paper, the result obtained here is 
in fact larger than the numerical results show. 
The $G_{M\bar{M}}$ was also 
evaluated in many other  models \cite{MMmodels}. 
Depending upon the variant of the model,
the value of $G_{M\bar{M}}$ ranges from $10^{-9}G_F$ to $0.1G_F$.

The conversion probability $P(M\to\bar{M})$ is the quantity that is
measured in experiments. It is related to the constant $G_{M\bar{M}}$
in the following way \cite{FW61}
\begin{equation}
P(M\to\bar{M})\ =\ \frac{\delta^2}{2\Gamma_\mu^2},
\end{equation}
where
\begin{equation}
\frac{\delta}{2}\ =\ \langle\bar{M}|H|M\rangle\ =\ 
\frac{16G_{M\bar{M}}}{\pi a^3}
\end{equation}
is a transition matrix element between the muonium and antimuonium
states ($a$ is the radius of muonium atom) and $\Gamma_\mu$ is the total 
decay width of muon.

From the point of view of SM
extended by heavy neutrinos,
$M$-$\bar{M}$ conversion 
is not a good place to search for LFV. Roughly
speaking, the $M$-$\bar{M}$ amplitude is proportional to the square of the
nondiagonal $\mu$-$e$ mixing $\sum_iB_{\mu N_i}B^*_{eN_i}$, 
which is strongly constrained by the
measurements of processes $\mu\to e\gamma$, $\mu\to eee$ and $\mu\to e$
conversion. Amplitudes of the three processes depend approximatively  
linearly on the nondiagonal $\mu$-$e$ mixing. 
Therefore, if any of the experimental results of
the three processes is improved by a factor $a$, the experimental result for
$P(M\to\bar{M})$ has to be improved by the factor $a^2$ to be competitive in
finding LFV.

\subsection{Extension and correction of some previous results}
In this subsection some previous results on neutrinoless LFV processes 
evaluated within the frame of the $V$ model are extended and/or corrected.

The decays of $\tau$ lepton into three leptons 
were evaluated within the $V$ model 
in Ref. \cite{GGV92} without including 
terms with four $B_{lN}$-s. These terms 
were shown to dominate for large
heavy neutrino masses in SM extended by 
two additional heavy Majorana neutrinos \cite{IP_NPB}. 
In that model the $B_{lN}$-s are
completely determined by $s_L^{\nu_i}$ parameters and ratio of the
heavy neutrino masses. Here, the upper bounds of complete amplitudes 
are evaluated within the $V$ model, 
and used to find the upper bounds of the corresponding
BRs.

Neutrinoless LFV decays of the Z boson were 
studied in Ref. \cite{IP_NPB} in SM extended with
heavy Majorana neutrinos.
The expressions for loop functions are given in Appendix A
of that reference, and they 
are correct except for terms containing
\begin{equation}
\frac{\sqrt{w}}{\lambda_Z}\tan^{-1}
\bigg(\frac{\sqrt{w}}{\lambda_i+\lambda_j-\lambda_Z}\bigg),
\end{equation}
which should be replaced with the expression
\begin{eqnarray}
&&\theta(w)\frac{\sqrt{w}}{\lambda_Z}
\bigg[\tan^{-1}\bigg(\frac{\sqrt{w}}{\lambda_i+\lambda_j-\lambda_Z}\bigg)
+\pi\theta(\lambda_Z-\lambda_i-\lambda_j)\bigg]
\nonumber\\&& 
+\theta(-w)\frac{\sqrt{-w}}{\lambda_Z}
\bigg[\frac{1}{2}\ln\bigg|\frac{\lambda_Z-\lambda_i-\lambda_j+\sqrt{-w}}
{\lambda_Z-\lambda_i-\lambda_j-\sqrt{-w}}\bigg|
-i\pi\theta(\lambda_Z-\lambda_i-\lambda_j)\bigg].
\end{eqnarray}
The notation is the same as in Ref. \cite{IP_NPB}.
The theta function in the first square bracket was not taken into account
in the 
analysis in Ref. \cite{IP_NPB}. 
As it contributes only for the heavy neutrino
masses smaller than the 
$Z$-boson mass, the numerical results given there should not
change. For heavy neutrinos lighter than $Z$-boson mass, the theta
function assures the continuity of the loop functions in heavy neutrino
masses. 
Here, LFV decays of the $Z$ boson
are studied in the $V$ model. 
The terms containing the matrix elements $C^*_{N_iN_j}$, 
that exist only for heavy 
Majorana neutrinos, are neglected. 
In the $V$ model only the upper bounds of the 
$Z\to ll'$ amplitudes can be found. They are
found using the formalism of the section \ref{secThelim} .

The only three neutrinoless LFV processes 
that give additional constraints on 
$B_{lN}$-s, $\mu\to e\gamma$, $\mu\to eee$ and $\mu - e$
conversion, were examined in Ref. \cite{TBBJ95}. 
Their analysis has included
the "nondecoupling" effects of heavy neutrinos, 
has indicated that a generalization 
of  Appelquist--Carazzone theorem \cite{AC,SS} 
is valid for the $V$ model and has determined the limits on 
specific combinations of $B_{lN}$-s. The "proof" of 
the generalization of the Appelquist--Carazzone theorem is based on 
an introduction of a somewhat arbitrary 
maximal $SU(2)_L$-doublet mass term.
The amplitude 
they present for $\mu\to e$ conversion does not 
include the photon exchange and 
box contributions, and the amplitude for $\mu\to eee$ does not include
the photon exchange term. These terms are included here. 
Moreover, in their expression for 
$\mu\to 3e$ BR,
obtained in the limit of large heavy neutrino masses,
one has to make replacements 
${\cal F}_{e\mu}\to 2{\cal F}_{e\mu}$ and 
$\varepsilon_L\equiv -1/2+s_W^2\to -\varepsilon_L$,
(the notation of Ref. \cite{TBBJ95} is used).

The neutrinoless LFV decay of $\pi^0$ was studied in Ref. 
\cite{FI96} in extensions of SM with additional Majorana and 
additional Dirac neutrinos. The expressions for the extension with 
Majorana neutrinos is correct, but the expressions for the extension 
by Dirac neutrinos is not, because the terms existing only for
Majorana neutrinos were kept in the amplitude. The correct amplitude
is obtained neglecting the terms containing the loop function $H_Z$.
When this correction is made, 
the numerical results for the $\pi\to\mu e$ decay become
$\sim 25$ times smaller.

\section{On low energy neutrionoless LFV amplitudes and decay rates}
\label{secLENLFVADR}

\subsection{Loop functions included in LFV processes} 
In the lowest order of perturbation theory, 
amplitudes of neutrinoless LFV decays 
are built up from several building blocks (composite 
loop functions and tree-level
functions) which may
be denoted by the exchanged bosons, or by the type of
the Feynman diagram: $\gamma$, $Z$, $box$
(box containing only leptons, leptons and $u$ quarks, leptons and
$d$ quarks), $H$ and $W^+W^-$. 
All functions except the last one are combinations of loop functions
and $B_{lN}$-s \cite{IP_NPB,FI96,I_PRD96}.
$W^+W^-$ function is a tree-level function and it 
is strongly suppressed compared to the others \cite{I_PRD96}. 
$\gamma$, $Z$, $box$ and $H$ functions
comprise two-fermion currents.
In the $\gamma$, $Z$ and $H$ functions
only one of the fermion currents changes flavor,
while in box functions flavors may be changed in both fermion currents. The 
classification of the neutrinoless
LFV decays, given in Table I, is made 
according to the Feynman diagrams they contain and 
the approximations (physics) one has to use in finding
the corresponding amplitudes. The references cited in Table I refer
only
to the calculations of LFV 
processes in the extensions of SM by additional 
heavy neutrinos.

If the heavy neutrino masses are larger than a few hundred
$GeV$, the expressions for neutrinoless LFV decays simplify 
considerably. All amplitudes can approximately be 
expressed in terms of four combinations of
masses and $B_{lN_i}$-s, 
\begin{eqnarray}
\label{comb}
{\cal A}_{ll'}& =& \sum_{N_i}B_{lN_i}^*B_{l'N_i},
\nonumber\\
{\cal B}_{ll'}& =& \sum_{N_i}B_{lN_i}^*B_{l'N_i}\ln\lambda_{N_i},
\nonumber\\
{\cal C}_{ll'}& =& \sum_{N_iN_j}B_{lN_i}^*C_{N_iN_j}^*B_{l'N_i}
 \frac{\lambda_{N_i}\lambda_{N_j}}{\lambda_{N_i}-\lambda_{N_j}}
 \ln\frac{\lambda_{N_i}}{\lambda_{N_j}},
\nonumber\\
{\cal D}_{ll'l_1l_2}& =& \frac{1}{2}\sum_{N_iN_j}B_{lN_i}^*B_{l_2N_j}^*
(B_{l'N_i}B_{l_1N_j}+B_{l_1N_i}B_{l'N_j})
 \frac{\lambda_{N_i}\lambda_{N_j}}{\lambda_{N_i}-\lambda_{N_j}}
 \ln\frac{\lambda_{N_i}}{\lambda_{N_j}},
\end{eqnarray}
where $\lambda_{N_i}=m_{N_i}^2/m_W^2$. 
The building blocks mentioned above, expressed in terms of combinations 
(\ref{comb}), read
\begin{eqnarray}
\label{amp_msx}
G_\gamma^{ll'}& \approx& \frac{1}{2}{\cal A}_{ll'},
\nonumber\\
F_\gamma^{ll'}& \approx& -\frac{1}{6}{\cal B}_{ll'},
\nonumber\\
F_Z^{ll'}& \approx& -\frac{3}{2}{\cal B}_{ll'}-\frac{1}{2}{\cal
C}_{ll'},
\nonumber\\
F_{box}^{ll'l_1l_2}& \approx& -({\cal A}_{ll'}\delta_{l_1l_2}+
{\cal A}_{ll_1}\delta_{l'l_2})+\frac{1}{2}{\cal D}_{ll'l_1l_2},
\nonumber\\
F_{box}^{ll'u_au_b}& \approx& \bigg[-4\delta_{u_au_b}+
\bigg(-\frac{9}{4}\frac{\lambda_b}{1-\lambda_b}+
\frac{-\lambda_b^3+8\lambda_b^2-
16\lambda_b}{4(1-\lambda_b)}\ln\lambda_b
\bigg)V^*_{{u_a}b}V_{{u_b}b}\bigg]{\cal A}_{ll'}
\nonumber\\
&&+\bigg[\frac{\lambda_b}{4}
V^*_{{u_a}b}V_{{u_b}b}\bigg]{\cal B}_{ll'},
\nonumber\\
F_{box}^{ll'd_ad_b}& \approx& \bigg[-\delta_{d_ad_b}+
\sum_{u_i=c,t}\bigg(\frac{3}{4}\frac{\lambda_{u_i}}{1-\lambda_{u_i}}+
\frac{-\lambda_{u_i}^3+8\lambda_{u_i}^2-
4\lambda_{u_i}}{4(1-\lambda_{u_i})}\ln\lambda_{u_i}
\bigg)V_{{u_i}d_a}V^*_{{u_i}d_b}\bigg]{\cal A}_{ll'}
\nonumber\\
&&+\bigg[\sum_{u_i=c,t}\frac{\lambda_{u_i}}{4}
V_{{u_i}d_a}V^*_{{u_i}d_b}\bigg]{\cal B}_{ll'},
\nonumber\\
F_H^{ll'}& \approx& G_H^{ll'}\ \approx\ \frac{5}{8}{\cal A}_{ll'}+
\frac{\lambda_H}{4}{\cal B}_{ll'}+\frac{3}{4}{\cal C}_{ll'},
\nonumber\\
F_{W^+W^-}& \approx& \bigg(\sum_{i=1}^{n_G}V_{u_id_a}V^*_{u_id_b}\bigg)
{\cal A}_{ll'},
\end{eqnarray}
where $\lambda_x=m_x^2/m_W^2$, $x=b,t,H$.

For the important case of degenerate ($\lambda_{N_i}=\lambda_{N}$) 
and large heavy neutrino masses 
the functions (\ref{comb}) can be 
written in terms of parameters $s_L^{\nu_l}$ and $x^0_{ll'}$,
\begin{eqnarray}
\label{comb1}
{\cal A}_{ll'}& =& s_L^{\nu_l}s_L^{\nu_{l'}}x^0_{ll'},
\nonumber\\
{\cal B}_{ll'}& =& s_L^{\nu_l}s_L^{\nu_{l'}}x^0_{ll'}\ln\lambda_{N},
\nonumber\\
{\cal C}_{ll'}& =& s_L^{\nu_l}s_L^{\nu_{l'}}
 \sum_{i=1}^{n_G}(s_L^{\nu_i})^2x^0_{ll_i}x^0_{l_il'}\lambda_{N},     
\nonumber\\
{\cal D}_{ll'l_1l_2}& =&
\frac{1}{2}s_L^{\nu_l}s_L^{\nu_{l'}}s_L^{\nu_{l_1}}s_L^{\nu_{l_2}}
(x^0_{ll'}x^0_{l_1l_2}+x^0_{ll_1}x^0_{l'l_2})\lambda_{N}.
\end{eqnarray}
It is convenient to introduce four 
combinations of $B_{lN}$-s, 
heavy neutrino masses, $\lambda_N^{PUB}$ and upper bound values for 
$s_L^{\nu_l}$ parameters (\ref{sL}), denoted by $\tilde{s}_L^{\nu_l}$:
\begin{eqnarray}
x_{ll'}& =& {\cal A}_{ll'}
 \Big(\tilde{s}_L^{\nu_l}\tilde{s}_L^{\nu_{l'}}\Big)^{-1},
\nonumber\\
z_{ll'}& =& {\cal B}_{ll'}
 \Big(\tilde{s}_L^{\nu_l}\tilde{s}_L^{\nu_{l'}}
 \ln\lambda_{N}^{PUB}\Big)^{-1},
\nonumber\\
y_{ll'}& =& {\cal C}_{ll'}
 \Big(\tilde{s}_L^{\nu_l}\tilde{s}_L^{\nu_{l'}}
 \sum_{i=1}^{n_G}(\tilde{s}_L^{\nu_i})^2\lambda_{N}^{PUB}\Big)^{-1},
\nonumber\\
y_{ll'l_1l_2}& =& {\cal D}_{ll'l_1l_2}
 \Big(\tilde{s}_L^{\nu_l}\tilde{s}_L^{\nu_{l'}}
 \tilde{s}_L^{\nu_{l_1}}\tilde{s}_L^{\nu_{l_2}}
 \lambda_{N}^{PUB}\big)^{-1}
\end{eqnarray}
Any of these combinations is always smaller than one.

Here, few comments are in order. First, it is obvious that 
$|{\cal D}_{ll'l_1l_2}|\leq|{\cal C}_{ll'}|$ 
(the relation is also valid for 
large, nondegenerate heavy neutrino masses). Second, 
for degenerate neutrino masses, the function
${\cal C}_{ll'}$  becomes larger than the 
functions ${\cal A}_{ll'}$ and ${\cal B}_{ll'}$ if
\begin{equation}
\lambda_N\stackrel{\displaystyle >}{\sim}
\frac{1}{\sum_{i=1}^{n_G}(s_L^{\nu_i})^2}
\qquad
\mbox{and}
\qquad
\lambda_N\stackrel{\displaystyle >}{\sim}
\frac{\ln\lambda_N}{\sum_{i=1}^{n_G}(s_L^{\nu_i})^2},
\end{equation}
respectively. The dominance of the functions with quadratic mass 
dependence of the amplitude leads to the transient, 
so called "nondecoupling 
behaviour" of amplitudes. As explained in section \ref{secUBPFLFVA},
decoupling follows from PUB inequalities (\ref{PUB}).
A typical mass value for which the quadratic terms become larger
than the logarithmic terms is $m_N\sim 1500\ GeV$ for $s_L^{\nu_l}$ values of
the order of the present experimantal bounds (\ref{sL}).
Third, at the maximal $\lambda_N$ value permitted by the 
PUB ($\lambda_N^{PUB}$), 
the function ${\cal C}_{ll'}$ depends
essentially only on two diagonal mixing parameters, 
$s_L^{\nu_l}$ and $s_L^{\nu_{l'}}$, 
\begin{equation}
{\cal C}_{ll'}(\lambda_N^{PUB})\ =\ 
\frac{4n_R\sum_i (s_L^{\nu_i})^2x^0_{ll_i}x^0_{l_il'}}
{\alpha_W\sum_i(s_L^{\nu_i})^2}
s_L^{\nu_l}s_L^{\nu_{l'}}
\ \stackrel{\displaystyle <}{\sim}\ 
\frac{4n_R}{\alpha_W}s_L^{\nu_l}s_L^{\nu_{l'}}x_{ll'}
\ =\ \frac{4n_R}{\alpha_W}{\cal A}_{ll'}.
\end{equation}
Therefore, at $m_N=m_N^{PUB}$ 
all amplitudes depend essentially only on $s_L^{\nu_l}$ and
$s_L^{\nu_{l'}}$.
If both the 
logarithmic and quadratic mass terms are present in LFV amplitude, at
$m_N^{PUB}$ logarithmic terms contribute up to $\sim 10\%$ of the total 
amplitude. Fourth, if the $t$ quark contribution is
multiplied by small CKM matrix elements, 
box amplitudes may have large contribution coming from
$c$ quark in the loop expressions. 
For instance, in the processes $\tau\to eP^0/\mu P^0$
$c$ quark contribution to the amplitude is $\sim 13\%$.
Fifth, the processes containing only the function 
${\cal A}_{ll'}$ are most suitable for obtaining new information 
on $B_{lN_i}$ parameters, because they are almost independent of heavy
neutrino masses. 
Sixth, for degenerate heavy neutrinos the dependence 
of LFV amplitudes on LCKM matrix elements appears only through six
sums $\sum_iB_{lN_i}B^*_{l'N_i}$, $l\neq l'$ and $\sum_i|B_{lN_i}|^2$
(diagonal and nondiagonal mixings). 
Writing the sums in terms
of $s_L^{\nu_l}$-s and $x^0_{ll'}$-s, one can easily show that 
if some LFV amplitude tends to zero for $s_L^{\nu_l}\to 0$,
then the amplitude tends to zero for $x^0_{ll'}\to 0$, $l\neq l'$, too.
(Stricktly speaking reduction of a parameter $s_L^{\nu_l}$ 
by factor $a$ is equivalent to
the reduction $x^0_{ll}\to a^2x^0_{ll}$ and $x^0_{ll'}\to ax^0_{ll'}$, $l\neq
l'$, but, by definition, $x^0_{ll}=1$.)
This analysis shows that LFV
amplitudes may be reduced without changing the diagonal mixing parameters 
$s_L^{\nu_l}$. It also indicates that the absolute values of LFV 
amplitudes may attain any value between zero and the upper bound value.
Therefore,
LFV processes are unsuitable for finding the limits on the 
diagonal mixing parameters $s_L^{\nu_l}$.

\subsection{Approximative expressions for BRs 
in the large-mass limit and relations between them}
Keeping only the leading terms in the large-mass limit of heavy
neutrinos, the expressions for BRs 
of neutrinoless LFV decays may be expressed
in terms of the functions (\ref{comb}). In the following these expressions
are listed. The definitions of unknown quantities are given below the 
list.
\begin{equation}
\label{llgam}
B(l\to l'\gamma)\ \approx\ 
\frac{\alpha_W^3 s_W^2}{2^{10}\pi^2}\frac{m_l^5}{M_W^4\Gamma_l}\,
|{\cal A}_{ll'}|^2; 
\end{equation}
\begin{eqnarray}
\label{llll}
B(l^-\to l'^-l_1^-l_2^+,l_1=l_2\neq l')& \approx& 
\frac{\alpha_W^4}{3\times 2^{15}\pi^3}\frac{m_l^5}{M_W^4\Gamma_l}
\Big(|{\cal D}_{ll'l_1l_1}-(1-2s_W^2){\cal C}_{ll'}|^2
+|2s_W^2{\cal C}_{ll'}|^2\Big),
\nonumber\\
B(l^-\to l'^-l_1^-l_2^+,l'=l_1=l_2)& \approx&
\frac{\alpha_W^4}{3\times 2^{16}\pi^3}\frac{m_l^5}{M_W^4\Gamma_l}
\Big(|{\cal D}_{ll'l'l'}-2(1-2s_W^2){\cal C}_{ll'}|^2
+\frac{1}{2}|4s_W^2{\cal C}_{ll'}|^2\Big),
\nonumber\\
B(l^-\to l'^-l_1^-l_2^+,l_2\neq l',l_1)& \approx& 
\frac{\alpha_W^4}{3\times 2^{16}\pi^3}\frac{m_l^5}{M_W^4\Gamma_l}\,
|{\cal D}_{ll'l_1l_2}|^2;
\end{eqnarray}\\

\begin{equation}
\label{Zll}
B(Z\to l^-l'^++l^+l'^-)\ \approx\ \frac{\alpha_W^3}{3\times 2^8c_W^3}
\frac{M_W}{\Gamma_Z}|{\cal C}_{ll'}|^2;
\end{equation}\\

\begin{equation}
\label{mueconv}
R(\mu^-\, \mbox{Ti}\to\ e^-\, \mbox{Ti})\ \approx\ 
\frac{\alpha_W^4\alpha_{em}^3}{2^{10}\pi^2}
\frac{Z_{eff}^4}{Z}|F(-m_\mu^2)|^2Q_W^2
\frac{m_\mu^5}{M_W^4\Gamma_{capture}}\,|{\cal C}_{\mu e}|^2;
\end{equation}\\

\begin{equation}
\label{MMconv}
|G_{M\bar{M}}|\ \approx\ \frac{\alpha_W^2}{2^5M_W^2}\,
|{\cal D}_{\mu e e \mu}|;
\end{equation}\\

\begin{eqnarray}
\label{taulPcqf}
B(\tau\to lP^0,\mbox{cqf})& \approx& 
\frac{\alpha_W^4(\alpha_{P^0}^Z)^2}{2^{13}\pi}
\bigg(1-\frac{m_{P^0}^2}{m_\tau^2}\bigg)^2
\frac{m_\tau^3f_{P^0}^2}{M_W^4\Gamma_\tau}\,|{\cal C}_{\tau l}|^2,
\\
\label{taulVcqf}
B(\tau\to lV^0,\mbox{cqf})& \approx& 
\frac{\alpha_W^4(\alpha_{V^0}^Z)^2}{2^{13}\pi\gamma_{V^0}^2}
\bigg(1-\frac{m_{V^0}^2}{m_\tau^2}\bigg)^2
\bigg(1+2\frac{m_{V^0}^2}{m_\tau^2}\bigg)
\frac{m_\tau^3m_{V^0}^2}{M_W^4\Gamma_\tau}\,|{\cal C}_{\tau l}|^2,
\\
\label{taulPncqf}
B(\tau\to lP^0,\mbox{ncqf})& \approx& 
\frac{\alpha_W^4(\alpha_{P^0}^{box,ds})^2}{2^{11}\pi}
\bigg(1-\frac{m_{P^0}^2}{m_\tau^2}\bigg)^2
\frac{m_\tau^3f_{P^0}^2}{M_W^4\Gamma_\tau}\,|F_{box}^{\tau lds}|^2,
\\
\label{taulVncqf}
B(\tau\to lV^0,\mbox{ncqf})& \approx& 
\frac{\alpha_W^4(\alpha_{V^0}^{box,ds})^2}{2^{11}\pi\gamma_{V^0}^2}
\bigg(1-\frac{m_{V^0}^2}{m_\tau^2}\bigg)^2
\bigg(1+2\frac{m_{V^0}^2}{m_\tau^2}\bigg)
\frac{m_\tau^3m_{V^0}^2}{M_W^4\Gamma_\tau}\,|F_{box}^{\tau lds}|^2;
\end{eqnarray}\\

\begin{eqnarray}
\label{Pemucqf}
B(P^0\to e\mu,\mbox{cqf})& \approx& 
\frac{\alpha_W^4(\alpha_{P^0}^Z)^2}{2^{12}\pi}
\bigg(1-\frac{m_\mu^2}{m_{P^0}^2}\bigg)^2
\frac{m_{P^0}m_\mu^2f_{P^0}^2}{M_W^4\Gamma_{P^0}}\,|{\cal C}_{\mu e}|^2,
\\
\label{Pemuncqf}
B(P^0\to e\mu,\mbox{ncqf})& \approx&
\frac{\alpha_W^4(\alpha_{P^0}^{box,ds})^2}{2^{10}\pi}
\bigg(1-\frac{m_\mu^2}{m_{P^0}^2}\bigg)^2
\frac{m_{P^0}m_\mu^2f_{P^0}^2}{M_W^4\Gamma_{P^0}}\,|F_{box}^{\mu eds}|^2,
\\
\label{HMll}
B({\cal H}^0\to ll')& \approx&
\frac{\alpha_W^4}{2^{10}\pi}
\bigg(1-\frac{m_l^2}{m_{{\cal H}^0}^2}\bigg)^2
\frac{m_{{\cal H}^0}m_l^2f_{{\cal H}^0}^2}{M_W^4\Gamma_{{\cal H}^0}}
\,|F_{box}^{ll'q_aQ}|^2;
\end{eqnarray}\\

\begin{eqnarray}
\label{taulP1P2cqf}
B(\tau\to lP_1P_2,\mbox{cqf})& \approx&
\frac{\alpha_W^4}{2^{16}\pi^3}
\frac{\int_{(m_1+m_2)^2}^{(m_\tau-m_l)^2}
dt\,\alpha\, |\sum_{V^0}p_{BW}^{V^0}(q)\alpha_{V^0}^ZC_{V^0P_1P_2}|^2}
{M_W^4m_\tau^3\Gamma_\tau}\,|{\cal C}_{\tau l}|^2,
\\
\label{taulP1P2ncqf}
B(\tau\to lP_1P_2,\mbox{ncqf})& \approx&
\frac{\alpha_W^4(\alpha_{K^{*0}}^{box,sd})^2|C_{K^{*0}P_1P_2}|^2}{2^{14}\pi^3}
\,|F_{box}^{\tau lsd}|^2
\nonumber\\&&
\times\frac{\int_{(m_1+m_2)^2}^{(m_\tau-m_l)^2}
dt\,\Big[\alpha
+2\frac{m_1^2-m_2^2}{m_{K^{*0}}^2}\beta 
+\Big(\frac{m_1^2-m_2^2}{m_{K^{*0}}^2}\Big)^2\gamma\Big]
\, |p_{BW}^{K^{*0}}(q)|^2}
{M_W^4m_\tau^3\Gamma_\tau},
\\
\label{taulP1P2cqfH}
B(\tau\to lP_1P_2,\mbox{cqfH})& \approx&
\frac{\alpha_W^4}{2^{16}\pi^3}
\frac{M_{HP_1P_2}^4\int_{(m_1+m_2)^2}^{(m_\tau-m_l)^2}dt\,
\iota}{M_H^4M_W^4m_\tau\Gamma_\tau}
\,\bigg|\frac{3}{2}{\cal C}_{\tau l}\bigg|^2;
\end{eqnarray}\\[.3cm]

\begin{eqnarray}
\label{P1P2mue}
B(K_W\to \pi\mu^\mp e^\pm)& \approx&
\frac{\alpha_W^4\tilde{c}_{K^{*0}K_W\pi}^2}
{2^{14}\pi^3}
\frac{\int^{(m_{P_1}-m_{P_2})^2}_{(m_\mu-m_e)^2}dt\,
[A_{++}f_+^2+A_{+-}f_+f_-+A_{--}f_-^2]}{M_W^4m_{K}^3\Gamma_{K_W}}
\,|F_{box}^{\mu lsd}|^2;
\end{eqnarray}\\[.3cm]

\begin{eqnarray}
\label{HMPll}
B({\cal H}_a\to Pl'^-l^+)& \approx&
\frac{\alpha_W^4(N_{P}^{q_a})^2}{2^{13}\pi^3}
\frac{\int_{(m_l+m_{l'})^2}^{(m_{{\cal H}_a}-m_P)^2}dt\,
[a_P^2Z_{P1}+a_Pb_PZ_{P2}+b_P^2Z_{P3}]}
{M_W^4m_{{\cal H}_a}^3\Gamma_{{\cal H}_a}}
\,|F_{box}^{ll'Qq_a}|^2,
\\
\label{HMVll}
B({\cal H}_a\to Vl'^-l^+)& \approx&
\frac{\alpha_W^4(N_{V}^{q_a})^2}{2^{12}\pi^3}
\,|F_{box}^{ll'Qq_a}|^2\frac{1}{M_W^4m_{{\cal H}_a}^3\Gamma_{{\cal H}_a}}
\int_{(m_l+m_{l'})^2}^{(m_{{\cal H}_a}-m_V)^2}dt\,
[a_V^2Z_{V1}+b_V^2Z_{V2}
\nonumber\\&&
+c_V^2Z_{V3}+d_V^2Z_{V4}+
a_Vc_VZ_{V5}+b_Vc_VZ_{V6}+b_Vd_VZ_{V7}+c_Vd_VZ_{V8}];
\end{eqnarray}\\[.3cm]

\begin{eqnarray}
\label{BBemu}
B(B\to B'e\mu)& \approx&
\frac{\alpha_W^4}{2^{10}\pi^3}\,|F_{box}^{\mu eds}|^2
\frac{1}{M_W^4m_B^3\Gamma_B}
\int_{(m_\mu+m_e)^2}^{(m_B-m_{B'})^2}dt
[A_1(f_1^2+g_1^2)
+A_2(f_1^2-g_1^2)
\nonumber\\&&
+A_3(f_1g_1)+A_4(g_1g_3)+A_5(g_3^2)].
\end{eqnarray}\\

\noindent
All expressions are written in terms of products of dimensionless factors.
For the expressions containing the dominant term ${\cal C}_{ll'}$, the error
one makes by keeping only the dominant term is of the order 
$\stackrel{\displaystyle >}{\sim} 20\%$,
because the term ${\cal C}_{ll'}$ is always accompanied 
with the ${\cal B}_{ll'}$ term giving $\sim 10\%$ contribution to the
amplitude at $m_N^{PUB}$. 
Following abbreviations are used: $s_W=\sin\theta_W$, $c_W=\cos\theta_W$
($\theta_W$ is Weinberg's angle), $s_P=\sin\theta_P$, $c_P=\cos\theta_P$
($\theta_P$ is the mixing angle for psudoscalar nonet states) and
$s_V=\sin\theta_P$, $c_V=\cos\theta_P$
($\theta_V$ is the mixing angle for vector nonet states). 
In Eq. (\ref{mueconv}) $\alpha_{em}=1/137$ is the fine structure constant,
$Z$ is atomic number (for $^{48}_{22}Ti$ 
$Z=22$, $N=A-Z=26$), $Z_{eff}=17.6$
\cite{COKFV93,BNT93,FW62} is the effective atomic number of $Ti$ \cite{JCS59}, 
$F(-m_\mu^2)=0.54$ 
is its nuclear form factor \cite{SMC70,DFMRL74}
at momentum transfer
$q^2\approx -m_\mu^2$ \cite{COKFV93}, $Q_W=Z(1-4s_W^2)-N$ is the
coherent nuclear charge associated with coupling of $Z$ boson to 
nucleus \cite{BNT93}
and $\Gamma_{capture}$ is the capture rate for negative muons on Ti
\cite{BNT93,SMR87,MR97}. In Eqs. (\ref{taulPcqf}--\ref{HMll})
$f_P$ and $f_{\cal H}$ are decay constants of light and
heavy pseudoscalar mesons respectively, and $\gamma_{V^0}$ are constants
defining the decay constants for light vector mesons, 
$f_V=m_V/\gamma_V$. The normalizations used here are
\begin{eqnarray}
\label{cnorm}
{\cal A}^\mu_P(x)& =& if_P\partial^\mu P(x),
\nonumber\\
{\cal V}^\mu_V(x)& =& \left\{
\begin{array}{ll}
\frac{m_V^2}{\gamma_V}V^\mu(x) & \qquad \mbox{for light vector mesons,}\\
f_Vm_V V^\mu(x) & \qquad \mbox{for heavy vector mesons,}
\end{array}
\right.\ 
\end{eqnarray}
where ${\cal A}^\mu_{P}$ and ${\cal V}^\mu_{V}$ are the axial vector 
(vector current) with the same quark content as corresponding 
pseudoscalar meson $P(x)$ and vector meson $V(x)$ fields,
respectively. $\alpha_{P^0}^Z$, $\alpha_{V^0}^Z$,
$\alpha_{P^0}^{box,d_ad_b}$ and $\alpha_{V^0}^{box,d_ad_b}$ are
constants defining the meson content in vector and axial-vector 
quark currents contained in quark combinations in the $Z$- and $box$- amplitude.
They are
defined in Table II. The mass of the lighter lepton 
is neglected in the expressions (\ref{taulPcqf}--\ref{HMll}). 
As the composite loop function $F_{box}^{ll'q_aq_b}$ contains two
terms of approximatively equal magnitude, for brevity the expresions
of these BRs are not
written in terms of the functions (\ref{comb1}). The expressions
(\ref{taulP1P2cqf}--\ref{BBemu}) contain three body phase integrals.
The phase functions may be found in the following references:
$\alpha$, $\beta$, $\gamma$ and $\iota$ in Ref. \cite{I_PRD96}; 
$A_{++}$, $A_{+-}$ and $A_{--}$ in Ref. 
\cite{FI96}; $A_1$, $A_2$, $A_3$, $A_4$
and $A_5$ in Ref. \cite{FI96}. In Eq. (\ref{taulP1P2cqf}), 
$C_{V^0P_1P_2}=g_{\rho^0\pi^+\pi^-}c_{V^0P_1P_2}$ are 
constants stemming from the gauged chiral $U(3)_L\times U(3)_R/U(3)_V$
Lagrangian \cite{I_PRD96} (e.g. $c_{\rho^0\pi^+\pi^-}=1$), and 
\begin{equation}
p_{BW}^{V^0}(q)=\frac{1}{\gamma_{V^0}}\frac{m_{V^0}^2-i\Gamma_{V^0}m_{V^0}}
{m_{V^0}^2-i\Gamma_{V^0}m_{V^0}-q^2}\ \equiv\ \frac{1}{\gamma_{V^0}}
p_{BW}^{V^0,norm}(q)
\end{equation}
is the Breit-Wigner propagator for a vector meson $V^0$
multiplied by slightly modified expression $m_V^2/\gamma_V$. The modification of
the expression $m_V^2/\gamma_V$ is made to obtain 
$p_{BW}^{V^0,norm}(0)=1$ 
\cite{I_PRD96,FWW80ACS81,Mir9396,KPP8488}. The constant $g_{\rho^0\pi^+\pi^-}$
is equal to the $\rho$ self couplig constant $g_V$ from section 
\ref{secNLFVDOHM}. In Eq. (\ref{taulP1P2cqfH})
$M_{HP_1P_2}^2$ are mass parameters contained in the effective 
Higgs--meson Lagrangian \cite{I_PRD96},
\begin{eqnarray}
{\cal L}_{HMM}& =& \frac{g_W}{4M_W}\bigg[m_\pi^2((\pi^0)^2+2\pi^+\pi^-)
+2m_{K^+}^2K^+K^-+2m_{K^0}^2K^0\bar{K}^0
+\frac{m_{K^+}^2+m_{K^0}^2+m_\pi^2}{3}\eta_1^2
\nonumber\\&&
+\frac{2m_{K^+}^2+2m_{K^0}^2-m_\pi^2}{3}\eta_8^2
+\frac{2^{3/2}(2m_\pi^2-m_{K^+}^2-m_{K^0}^2)}{3}\eta_8\eta_1\bigg],
\end{eqnarray}
obtained by comparing the quark mass 
Lagrangian with the corresponding term in the chiral Lagrangian 
\cite{BBGCFG} (e.g.
$M_{H\pi^+\pi^-}^2=2m_\pi^2\equiv 2(2m_{\pi^+}^2+m_{\pi^0}^2)/3$). 
In Eq. (\ref{P1P2mue}) $\tilde{c}_{K^{*0}K_W\pi}=ac_{K^{*0}\bar{K}_S\pi}+
bc_{\bar{K}^{*0}K_S\bar{\pi}}$ ($K_W=a\bar{K}_S+bK_S$ is a weak kaon
eigenstate, and $K_S$ is a mass eigenstate). 
In Eq. (\ref{BBemu}) $f_1$, $g_1$ and $g_3$ are baryon form
factors. The other baryon form factors do not contribute, because they
belong to the second class currents, or give a contribution proportional to
the difference of baryon masses. 
The form factors $f_1$ and $g_1$ can be defined in terms of
two $SU(3)$ Clebsh-Gordan coeficients and two 
reduced matrix elements corresponding to symmetric and antisymmetric
octet $SU(3)$ representations. These reduced 
matrix elements are almost independent of 
momentum transfer and are usually identified with their value at zero
momentum transfer, $D$ and $F$.
The functions $g_1$ and $g_3$ are not independent, but
correlated through the Goldberger-Treiman relation \cite{FI96}.

All approximative expressions for the BRs
(\ref{llgam}-\ref{BBemu}), valid in the large mass limit, except
$B(l^-\to l'^-l_1^-l_2^+,l_1=l_2\neq l')$ and
$B(l^-\to l'^-l_1^-l_2^+,l'=l_1=l_2)$,
depend only on one of the 
functions (\ref{comb}). In the following, the smaller of 
two dominant functions ${\cal D}_{ll'l_1l_2}$ will be neglected
in the two exceptional expressions. The maximal error one makes in
the evaluation of BRs of the exceptional processes is $\sim 40\%$. 
With those approximations, the ratios of BRs 
having the same dominant function (\ref{comb}) 
become independent of the $V$-model parameters:\\

\noindent
a) BRs with $F_Z^{\mu e}$ (${\cal C}_{\mu e}$): 
\begin{eqnarray}
\label{FZmue}
&&R(\mu \mbox{Ti}\to e \mbox{Ti})\ :\ B(\mu\to ee^-e^+)\ :\ 
B(Z\to\mu^\mp e^\pm)\ :\ B(\pi^0\to\mu^\mp e^\pm)\ :\ 
B(\eta\to\mu^\mp e^\pm)
\nonumber\\
&&\ =\ 1\ :\ 5.60\times 10^{-2}\ :\ 3.77\times 10^{-2}\ :\ 
6.05\times 10^{-10}\ :\ 1.69\times 10^{-11},
\end{eqnarray}
b) BRs with $F_Z^{\tau l}$ (${\cal C}_{\tau l}$):
\begin{eqnarray}
\label{FZtaul}
&&B(Z\to\tau^\mp l^\pm)\ :\ B(\tau\to l\pi^0)\ :\ 
B(\tau\to l\rho^0)\ :\ B(\tau\to l\pi^+\pi^-)\ :\
B(\tau\to l\phi)
\nonumber\\
&&\ :\ B(\tau\to ll^-l^+)\ :\ B(\tau\to ll_1^-l_1^+)\ :\
B(\tau\to lK^+K^-)\ :\ B(\tau\to lK^0\bar{K}^0)\ :\
B(\tau\to l\eta')
\nonumber\\ 
&&\ :\ B(\tau\to l\eta)\ :\ B(\tau\to l\omega)\ :\
B(\tau\to l\eta\eta)\ :\ B(\tau\to l\pi^0\pi^0)
\nonumber\\
&&\ =\ 1\ :\ 3.40\times 10^{-1}\ :\ 3.17\times 10^{-1}\ :\ 
2.83\times 10^{-1}\ :\ 2.81\times 10^{-1}
\nonumber\\
&&\ :\ 2.64\times 10^{-1}\ :\ 1.64\times 10^{-1}\ :\ 
1.20\times 10^{-1}\ :\ 7.43\times 10^{-2}\ :\ 
6.15\times 10^{-2}
\nonumber\\
&&\ :\ 4.72\times 10^{-2}\ :\ 8.78\times 10^{-3}\ :\ 
4.34\times 10^{-12}\ :\ 5.50\times 10^{-13},
\end{eqnarray}
c) BRs with $F_{box}^{\mu esd}$ 
\begin{eqnarray}
\label{Fboxmuesd}
&&B(K_L\to\mu^\mp e^\pm)\ :\ B(K^+\to\pi^+\mu^\mp e^\pm)\ :\ 
B(\Sigma^+\to p\mu^\mp e^\pm)\ :\ B(\Xi^0\to\Lambda\mu^\mp e^\pm)
\nonumber\\
&&\ :\ B(\Lambda\to n\mu^\mp e^\pm)\ :\ B(\Xi^-\to\Sigma^-\mu^\mp e^\pm)
\ :\ B(\Xi^0\to\Sigma^0\mu^\mp e^\pm)\ :\ B(\Sigma^0\to n\mu^\mp e^\pm)
\nonumber\\
&&\ =\ 1\ :\ 3.01\times 10^{-2}\ :\ 1.30\times 10^{-4}\ :\  
1.21\times 10^{-4}
\nonumber\\
&&\ :\ 8.66\times 10^{-5}\ :\ 6.40\times 10^{-7}\ :\ 4.07\times 10^{-7}\ :\ 
6.31\times 10^{-14},
\end{eqnarray}
d) BRs with $F_{box}^{\mu ebd}$
\begin{eqnarray}
\label{Fboxmuebd}
B(B^-\to\pi^-\mu^\mp e^\pm)\ :\ B(B^0\to\mu^\mp e^\pm)\ =\ 
1\ :\ 3.76\times 10^{-4},
\end{eqnarray}
e) BRs with $F_{box}^{\mu ebs}$
\begin{eqnarray}
\label{Fboxmuebs}
&&B(B^-\to K^{*-}\mu^\mp e^\pm)\ :\ 
B(\bar{B}^0\to K^{*0}\mu^\mp e^\pm)\ :\ 
B(\bar{B}_s^0\to\eta'\mu^\mp e^\pm)\ :\ 
B(\bar{B}_s^0\to\phi\mu^\mp e^\pm)
\nonumber\\
&&\ :\ B(B^-\to K^-\mu^\mp e^\pm)\ :\ 
B(\bar{B}^0\to\bar{K}^0\mu^\mp e^\pm)\ :\ 
B(\bar{B}_s^0\to\eta\mu^\mp e^\pm)\ :\
B(\bar{B}_s^0\to\mu^\mp e^\pm)
\nonumber\\
&&\ =\ 1\ :\ 9.34\times 10^{-1}\ :\ 
8.83\times 10^{-1}\ :\ 8.57\times 10^{-1}
\nonumber\\
&&\ :\ 7.92\times 10^{-1}\ :\ 7.47\times 10^{-1}\ :\ 
3.31\times 10^{-1}\ :\ 4.93\times 10^{-4},
\end{eqnarray}
f) BRs with $F_{box}^{\tau lds}$
\begin{eqnarray}
\label{Fboxtaulds}
&&B(\tau\to e\pi^+K^-)\ :\ B(\tau\to eK^{*0})\ :\ B(\tau\to eK^0)
\ =\ 1\ :\ 7.32\times 10^{-1}\ :\ 2.99\times 10^{-1},
\end{eqnarray}
g) BRs with $F_{box}^{\tau lbd}$
\begin{eqnarray}
\label{Fboxtaulbd}
&&B(B^-\to\pi^-\tau^\mp e^\pm)\ :\ 
B(\bar{B}^0\to\tau^\mp e^\pm)
\ =\ 1\ :\ 1.14\times 10^{-1},
\end{eqnarray}
h) BRs with $F_{box}^{\tau lbs}$
\begin{eqnarray}
\label{Fboxtaulbs}
&&B(B^-\to K^{*-}\tau^\mp e^\pm)\ :\ 
B(\bar{B}^0\to\bar{K}^{*0}\tau^\mp e^\pm)\ :\ 
B(\bar{B}_s^0\to\phi\tau^\mp e^\pm)\ :\
B(\bar{B}_s^0\to\eta'\tau^\mp e^\pm)
\nonumber\\
&&\ :\ B(B^-\to K^-\tau^\mp e^\pm)\ :\
B(\bar{B}^0\to\bar{K}^0\tau^\mp e^\pm)\ :\
B(\bar{B}_s^0\to\eta\tau^\mp e^\pm)\ :\
B(\bar{B}_s^0\to\tau^\mp e^\pm)
\nonumber\\
&&\ =\ 1\ :\ 9.37\times 10^{-1}\ :\ 
8.10\times 10^{-1}\ :\ 6.57\times 10^{-1}
\nonumber\\
&&\ :\ 6.54\times 10^{-1}\ :\ 6.16\times 10^{-1}\ :\ 
2.76\times 10^{-1}\ :\ 1.63\times 10^{-1}.
\end{eqnarray}
For each group of BRs the BRs are lined up in the descending order. 
For instance, $\mu\to e$ conversion is the most suitable for finding 
LFV in the group containing the composite loop function $F_Z^{\mu e}$.
The position in the group depends on the coupling constants, phase factors
and the total decay rate of the decaying particle. For instance, the
BRs for $\tau\to lP_1P_2$ processes containing $Z$-boson amplitude are 
$\sim 10^{12}$ times larger than BRs 
of the processes $\tau\to lP_1P_2$ containing
only Higgs amplitude, because of the small Higgs-meson couplings,
although
the dominant composite loop functions are essentially the same.
The ratios (\ref{FZmue}-\ref{Fboxtaulbs}) are given 
for measured processes and LFV processes that
have not been studied in models with additional heavy neutrinos 
before. The $l\to l'\gamma$ decays 
are not included in the above ratios, because each process $l\to l'\gamma$
forms a group for itself, depending only on the function ${\cal A}_{ll'}$.
The numerical results for the ratios of BRs 
agree quite well with the exact ratios for degenerate heavy neutrino masses
obtained at $m_N=m_N^{PUB}$.
That allows one to consider only one of the decays of each group when 
comparing theoretical and experimental results.

Besides the ratios of BRs having the same dominant function 
(\ref{comb}), it is usefull to have relations that relate  
BRs of different groups of decays. These relations generally 
depend on the matrix elements of $B$ matrix and CKM matrix.
For instance,
\begin{eqnarray}
\label{R1}
&&B(Z\to\tau^\mp e^\pm)\ :\ 
B(Z\to\tau^\mp\mu^\pm)\ :\ 
B(Z\to\mu^\mp e^\pm)
\nonumber\\
&&\ =\ |F_Z^{\tau e}|^2\ :\ |F_Z^{\tau \mu}|^2\ :\ |F_Z^{\mu e}|^2,
\nonumber\\ 
&&B(\tau\to eP^0)\ :\ B(\tau\to\mu P^0)\ :\ B(P^0\to e^\mp \mu^\pm)
\nonumber\\
&&\ =\ \left\{\begin{array}{l}
|F_Z^{\tau e}|^2\ :\ |F_Z^{\tau \mu}|^2\ :\ 
\frac{1.45\times 10^9\, s^{-1}}{\Gamma_{P^0}}
\frac{m_{P^0}}{m_\mu}\frac{(1-m_\mu^2/m_{P^0}^2)^2}
{(1-m_{P^0}^2/m_\tau^2)^2}\: |F_Z^{\mu e}|^2\qquad \mbox{for cqf},\\
|F_{box}^{\tau eds}|^2\ :\ |F_{box}^{\tau\mu ds}|^2\ :\ 
\frac{1.45\times 10^9\, s^{-1}}{\Gamma_{P^0}}
\frac{m_{P^0}}{m_\mu}\frac{(1-m_\mu^2/m_{P^0}^2)^2}
{(1-m_{P^0}^2/m_\tau^2)^2}\: |F_{box}^{\mu eds}|^2\qquad  \mbox{for ncqf},
\end{array}\right.
\nonumber\\
&&B(\tau\to e\gamma)\ :\ B(\tau\to\mu\gamma)\ :\ B(\mu\to e\gamma)
\nonumber\\
&&\ =\ |B_{\tau N}^*B_{eN}|^2\ :\ 
|B_{\tau N}^*B_{\mu N}|^2)^2\ :\ 5.63|B_{\mu N}^*B_{eN}|^2,
\nonumber\\
&&B(\tau\to ee^\mp e^\pm)\ :\ B(\tau\to\mu\mu^\mp \mu^\pm)\ :\ 
B(\mu\to ee^\mp e^\pm)
\nonumber\\
&&\ \approx\ |F_Z^{\tau e}|^2\ :\ |F_Z^{\tau \mu}|^2\ :\ 
5.63|F_Z^{\mu e}|^2,
\nonumber\\
&&B(\tau\to eK^+\pi^-)\ :\ B(\tau\to\mu K^+\pi^-)\ :\ 
B(K^-\to\pi^-\mu^\mp e^\pm)
\nonumber\\
&&\ =\ 
|F_{box}^{\tau eds}|^2\ :\ 0.983|F_{box}^{\tau\mu ds}|^2\ :\ 
6.82|F_{box}^{\mu eds}|^2,
\nonumber\\
&&B(B_i\to l_1^\mp l_2^\pm)\ :\ B(B_j\to l_3^\mp l_4^\pm)\ =\ 
|F_{box}^{l_2l_1q_ib}|^2\ :\ |F_{box}^{l_4l_3q_jb}|^2\qquad 
q_i,q_j=u,d,s.
\end{eqnarray}

BRs of processes having only the 
logarithmic dependence on mass are several orders
of magnitude smaller than BRs
containing the quadratic mass dependent terms. In the processes
containing quarks in the final state, the presence of 
small CKM matrix elements 
additionally reinforces this difference.
E.g., at $m_N^{PUB}$
\begin{eqnarray}
\label{R2}
&&B(l\to l'\gamma)\ :\ B(l\to l'l_1l_2)\ 
\stackrel{\displaystyle <}{\sim}\ 10^{-2},
\nonumber\\
&&B(\tau^-\to e^-K^0)\: :\: B(\tau^-\to e^-\pi^0)
\: \approx\: B(\tau^-\to e^-\pi^+K^-)\: :\: 
B(\tau^-\to e^-\pi^-\pi^+)\: 
\stackrel{\displaystyle <}{\sim}\:  10^{-9}.
\end{eqnarray}
Between the LFV decays having the box contribution only, the $B$-meson 
decays have the largest CKM matrix elements. For that reason they might
be the most suitable box-dominated processes
for finding LFV in the future $B$-factories.

\subsection{Numerical results, comparison with experiment and
discussion}

In this subsection 
the experimental upper bounds for the measured neutrinoless LFV BRs
are compared with the theoretical upper bounds obtained in the 
$V$ model. For some interesting
unmeasured processes, the theoretical upper bounds are given, too.
The results are discussed. The limit on the nondiagonal $\mu$-$e$ mixing
is updated.
The decoupling of very heavy neutrinos is shown
explicitely. The possible error one can make using the upper bound
procedure given in the section \ref{secUBPFLFVA} is estimated.

Theoretical results depend on the $V$-model parameters: 
"diagonal" mixings $s_L^{\nu_l}$, phases of $B_{lN}$-s
and heavy neutrino masses. 
The parameters $s_L^{\nu_l}$ must satisfy the experimental upper
bounds (\ref{sL}), the heavy neutrinos are bound by the PUB inequalities
(\ref{PUB}) and (\ref{PUB_PV}), while the phases of $B_{lN}$ matrices are
undetermined. 
The numerical results are largest for degenerate neutrino masses
at maximal 
values of $s_L^{\nu_l}$ parameters 
and maximal neutrino mass, $m_N^{PUB}$. 
For degenerate heavy neutrino
masses the phase dependence of $B_{lN}$ matrices is contained in the
parameters $x_{ll'}$. 

The numerical values for BRs and $G_{M\bar{M}}$ depend on a number 
of "SM" particle properties, too: 
decay rates of particles, masses of the
particles included in the decays, CKM matrix elements, decay constants
of mesons, quark masses included in loops, mixing angles, various 
couplings, etc. Almost all these quantities are taken from Ref.
\cite{PDG98}, or derived from the 
data given there. For instance, masses of the 
$u$, $d$, $s$, $c$ and $b$ quarks are taken to be equal to the average of 
the upper and lower bound values. The CKM matrix elements are derived in the 
same way. The $t$ quark mass is set to be equal 
to the experimental value obtained from 
the direct observations of $t$ quark. For pseudoscalar meson decay constants
of light mesons we took the values partly from Ref. \cite{PDG98} and
partly from  Ref. \cite{PDG94},
\begin{eqnarray}
&&f_{\pi^+}\ =\ 130.7\ MeV,\quad   f_{K^+}\ =\ 159.8\ MeV,\quad 
f_{\pi^0}\ =\ 119\ MeV,\quad 
\nonumber\\
&&f_\eta\ =\ 131\ MeV\quad \mbox{and}
\quad f_{\eta'}\ =\ 118\ MeV.
\end{eqnarray}
Due to the isospin symmetry, $f_{K^0}=f_{\bar{K}^0}=f_{K^\pm}$. 
The constants $\gamma_{V^0}$, defining the decay constants of light vector
mesons, are extracted from the $V^0\to e^+e^-$ decay rates 
\begin{equation}
\gamma_{\rho^0}\ =\ 2.518,\quad
\gamma_{\phi}\ =\ 2.933,\quad
\gamma_\omega\ =\ 3.116,
\end{equation}
or estimated 
using the SU(3) octet symmetry, 
$\gamma_{K^{*0}}=\gamma_{\rho^0}$. For all decay constants of
$D$ and $D^*$ mesons, the conservative value $200\ MeV$ is taken. 
The decay constants of $B$ and $B^*$ mesons are derived using the scaling
law for decay constants derived from HQET,
\begin{equation}
f_{\cal  H}\ \sim\ m_{\cal  H}^{-1/2}.
\end{equation}
The weak fine-structure constant is defined as
$\alpha_W=\alpha_{em}/\sin^2\theta_W$, with $\cos\theta_W=M_W/M_Z$.
The $\rho-\pi-\pi$ coupling constant (which is equal to the $\rho$ meson
self coupling constant) is derived from the $\rho\to 2\pi$
coupling width. Other vector-meson--pseudoscalar-meson couplings of light
mesons are fixed by one of the chiral model Lagrangians 
\cite{BKY88,I_PRD96}. The mixing of the vector-meson nonet states is
determined from the quadratic Gell-Mann--Okubo mass formula, 
$\theta_V=39.3^\circ$. The mixing of the pseudoscalar-meson nonet
states is extracted from the $e^+e^-\to e^+e^-\gamma\gamma^*\to e^+e^-
(P\to\gamma\gamma)$ experiments \cite{thetaP}, $\theta_P=-23^\circ$. 
The only "SM" parameters that are not firmly established are 
"HQET+CHPT" parameters 
describing the semileptonic LFV decays of the $B$-mesons, $g$, $\beta$,
$\lambda$, $\alpha_1$ and $\alpha_2$ (see section \ref{secNLFVDOHM}).
The corresponding parameters for $D$-mesons have been determined 
by fitting the theory to the experimental values of the semileptonic decays
of $D$ mesons \cite{BFOP97,FPS98,F99}. The $B$-meson parameters 
$\lambda$, $\alpha_1$ and $\alpha_2$ may 
be derived from the $D$-meson parameters 
from the scaling laws for the vector and axial current
\cite{Cas97}. The parameter $g$ is independent of a heavy quark mass,
and the value of parameter $\beta$ is consistent with zero. The best
$B$-meson parameters obtained using the above procedure are \cite{F99},
\begin{eqnarray}
g& =& 0.2,\quad \beta\ =\ 0,\quad 
\lambda\ =\ -0.34\ GeV^{-1},\quad
\nonumber\\
\alpha_1& =& -0.13\ GeV^{1/2},\quad \alpha_2\ =\ -0.36\ GeV^{1/2}.
\end{eqnarray}
That way, all parameters are defined.

For measured
processes, the experimental and theoretical upper bounds of
the exact BRs are compared in Table IIIa. 
For some interesting processes that have not been
measured, the 
theoretical upper bounds are given in Table IIIb. 
In the both tables, the numerical part 
of the theoretical results 
is evaluated for degenerate heavy neutrino masses and the
maximal heavy neutrino mass permited by PUB, maximal $s_L^{\nu_l}$ values
and neglecting the $B_{lN}$ phases. The factors
$x_{ll'}$, $y_{ll'}$ and $y_{ll'l_1l_2}$ and $z_{ll'}$ 
describe the deviation 
of BRs from these values, when the model parameters assume other 
values.
The factors $y_{ll'}$
and $y_{ll'l_1l_2}$ give only the behavior of the dominant, 
$m_N^2$-dependent term, on the model parameters.
For $m_N$ values 
for which the terms quadratic in $B_{lN}$
matrices begin to dominate ($m_N\sim 1000-1500\ GeV$), 
the $z_{ll'}$ terms begin to dominate.

Comparing the theoretical upper bounds for the processes of the same 
type with different leptons in the initial and final state, one can see
that they are often comparable in magnitude. For instance, upper bounds
for BRs
of the processes $l\to l'\gamma$, $l\to l'l_1l_2$ and $Z\to ll'$ are 
of the order $\sim 10^8$, $\sim 10^{-6}$ and $\sim 10^{-6}$, respectively.
For that reason, the muon LFV processes which have been measured with
the greatest precision,
are the most attractive for finding LFV. A process with weaker
experimental bounds may be interesting only if the parameter(s)
$x^0_{ll'}$ for that process is (are) large. 

If, for a specific process,
the theoretical upper bound is larger than the experimental one, 
then the process gives
the better bound on a specific combination of $B_{lN}$-s
than the limit (\ref{sL}). The processes for which this ratio is larger
than one are $\mu\to e\gamma$, $\mu\to eee$, $\mu Ti\to eTi$, $\tau\to
e\rho^0$, $\tau\to e\pi^+\pi^-$ and $Z\to e\tau$. For the last three
processes the ratio is very close to one. As their amplitudes are
dominated by $m_N^2$ part of the amplitude, the new limits on $B_{lN}$ 
combinatons contain $m_N^2$ mass dependence, too, and therefore are
uninteresting.
For the first
three processes the ratio is much larger than one, and they do give
new limits on specific combinations of $B_{lN}$-s
as shown in Ref. \cite{TBBJ95}. Since that paper was published, 
the limits on $\mu\to e\gamma$
and $\mu Ti\to eTi$ improved by factors 
$1.3$ ($4.1$ \cite{MDpk99}) and $7$,
respectively. The first of them gives a new limit 
on nondiagonal $\mu$-$e$ mixing,
\begin{equation}
\label{BeBmu2}
\sum_{i=1}^{n_G}B_{\mu N_i}B^*_{eN_i}\leq 2.15\times 10^{-4}
\ (1.19\times 10^{-4}).
\end{equation}
To obtain the limit on the nondiagonal $\tau$-$e$ and $\tau$-$\mu$
mixings, the present experimental sensitivities of $\tau\to l$ decays
should improve by two orders of magnitude.
It is interesting that the $\mu Ti\to eTi$ conversion also gives very
good mass independent limit on the sum $\sum_{i=1}^{n_G}B_{\mu
N_i}B^*_{eN_i}$. 
Namely,
$\mu Ti\to eTi$ amplitude contains mass independent part coming from
the photon exchange. If the terms in the $\mu Ti\to eTi$ amplitude do
not cancel completely, one can make an estimate of the sum  
by attributing the whole amplitude to the photon
excahnge part of the amplitude. That way  
one can only make a worse estimate of the sum. The limit one obtains 
that way reads
\begin{equation}
\label{BeBmu3}
\sum_{i=1}^{n_G}B_{\mu N_i}B^*_{eN_i}\leq 3.93\times 10^{-4}.
\end{equation}

For all processes whose amplitudes comprise only the box amplitude,
the theoretical upper bounds are several orders of magnitude
smaller than the experimental upper bounds. For the $K_L\to e^\mp\mu^\pm$
decay the ratio of theoretical and experimental upper bound is largest,
$1.58\times 10^{-3}x_{\mu e}^2$. As the present experimental limit
is $2\times 10^{-11}$ \cite{KPJ98}, its significant improvement cannot be
expected. Although the experimental upper bounds for
semileptonic LFV $B$-meson processes are weak, the corresponding
theoretical upper bounds are of the order $\sim 10^{-9}$. Therefore,
$B$-meson decays are interesting for finding LFV decays in the 
near future.

The recent Super-Kamiokande experiment shows there is a large mixing
between $\nu_\mu$  and some other light neutrino, very probably 
$\nu_\tau$. If the additional heavy neutrinos exist, this might suggest a
large "angle" parameter $x^0_{\tau\mu}$. Therefore, the Super-Kamiokande
result might be a sign to search for LFV among processes with tauon 
and muon in the final (and initial) state.

To estimate how large error one can make using the upper bound
procedure from section \ref{secUBPFLFVA}, the BRs for the processes 
$\mu Ti\to e Ti$, $Z\to \mu^\mp\tau^\pm$, $K_L\to e^\mp\mu^\pm$ and
$B^-\to K^{*-}\mu^\mp\tau^\pm$ are evaluated 
using both the upper bound procedure and the "realistic" $B_{lN}$-s 
(\ref{B_ex}). These processes are chosen because they have the
maximal BR within the group of processes with the same dominant composite
loop function. The first two of these processes contain $F_Z^{ll'}$ function 
and the last two 
contain $F_{box}^{ll'd_ad_b}$ function only. The BRs are evaluated 
for degenerate heavy neutrino masses and
two sets of $s_L^{\nu_l}$ and $x_{ll'}^0$ parameters for which the
maximal theoretical value for $B(\mu\, Ti\to e\, Ti)$ is equal to the
present experimental upper bound. The firts set is obtained from 
the "maximal set" ($s_L^{\nu_l}$-s from Eq. \ref{sL} and all $x_{ll'}=0$)
by replacing the maximal value for $(s_L^{\nu_e})^2$ with the value
$(s_L^{\nu_e})^2=4.29\times 10^{-10}=
7.1\times 10^{-3}\times(2.459\times 10^{-4})^2$. 
The second set is obtained from
the "maximal set" by putting $x_{\mu e}=x_{\tau e}=2.459\times 10^{-4}$. 
The first set is used together with upper bound procedure and with
"realistic" $B_{lN}$ matrix elements. The second can be applied only
within upper bound procedure, because the procedure with "realistic"
$B_{lN}$ matrix elements has fixed $x_{ll'}$ values.  In all
calculations 
$x_{\mu\tau}$ is kept to be equal one in accord with the
Super-Kamiokande results.
The BRs are presented in Fig. 1.
as functions of heavy neutrino mass. The figures illustrate 
the following properties of the BRs. First, for all $m_N$ values,
the upper bound procedure 
gives larger value than the "realistic" $B_{lN}$-s.
Second, while the BRs evaluated in the upper bound
procedure increase in the whole region of $m_N$ values permitted by PUB,
the BRs evaluated with the "realistic" $B_{lN}$-s may have
a maximum below the $m_N^{PUB}$. The maximum is a consequence of the 
mass dependence of the "realistic" $B_{lN}$-s. All BRs
of processes with $box$-amplitude
only have the maximum, but it can appear in the BRs having the 
$Z$-amplitude, too. 
Third, the by reduction of $x_{ll'}$-s one obtains the result which are
numerically equivalent to the results obtained by reduction of
$s_L^{\nu_l}$ parameters.
Fourth, a
strong cancelation of the amplitude terms may appear in the 
BRs evaluated with the "realistic" $B_{lN}$-s, as in the case 
of $\mu Ti\to e Ti$. Fifth, the error one can make 
in the evaluation of the
maximum of BRs using the upper bound procedure is 
$\stackrel{\displaystyle <}{\sim} 10$ for processes with the $box$-amplitude
only, and 
$\stackrel{\displaystyle <}{\sim} 100$ for processes with
$Z$-amplitudes. The flat behavior of $Z\to l^\mp l'^\pm$ at $m_N\sim
100\ GeV(\sim m_Z)$ is a consequence of treshold effects.

As shown in the section \ref{secThelim}, all heavy neutrino masses, except
one, can assume any value between zero and infinity. BRs should not assume
values larger than one in the whole parameter space permitted by the
model. The illustration of convergence and of good behavior of
branching ratios evaluated using the upper bound 
procedure and  "realistic" $B_{lN}$-s is given in Fig. 2. Branching
ratios are evaluated keeping two masses equal, while the third one is
assumeed to take very large variable values 
values. In Fig 2. the branching ratios 
for the same processes as in Fig 1. are given, but here as a 
function of ratio of the large mass ($m_{N_2}$) 
and mass which is kept constant ($m_{N_1}=m_{N_3}$).
Graphs in Fig 2.  show that the very 
heavy neutrinos decouple, and therefore, that the
nondecoupling of heavy neutrinos is only a transient effect. For the upper
bound procedure, the decoupling of the very heavy neutrino(s)
manifests as the equality of BR values
for degenerate heavy neutrinos and when some of masses tend to infinity,
while for "realistic" $B_{lN}$-s 
BR-s reduce in magnitude. Figure 2 also illustrates that the upper
bound procedure is very crude in the transient region where the 
upper bounds (\ref{PUB_APP}) and upper bound (\ref{PUB_PV}) are almost
equally 
effective as the second (\ref{PUB_APP}) bound. 
To show that with the proper choice of the parameters experimental
limits are always satisfied, 
in the first panel of Fig. 2 the additional BR curve is added, 
evaluated in the upper bound procedure for parameters for which 
the maximal BR value is smaller than the present experimental upper
bound for $R(\mu Ti\to e Ti)$. Only top of the curve is seen in the
figure. The curves obtained using the "realistic" $B_{lN}$-s are much
smoother than the curves following from the upper bound procedure.
Therefore, for nondegenerate heavy neutrinos good knowledge of the 
$B_{lN}$ matrix elements is necessary to obtain reasonable estimate of
the BR values.

\section{Conclusions}
\label{secC}
All low energy neutrinoless LFV processes are studied
in an extension of SM by heavy $SU(2)_L\times U(1)$
singlet Dirac neutrinos. 
The structure of amplitudes and relations between BRs 
are carefully analized. It is shown that, in principle, 
the neutrinoless LFV decays cannot give new limits on the "diagonal"
mixings $s_L^{\nu_l}$. The approximative 
expressions for all BRs are listed, keeping only the dominant terms
of the corresponding amplitudes in the large heavy-neutrino mass limit.
The approximative BRs are compared 
within the groups of processes with the same
dominant composite loop function, and within each group 
the experimentally most interesting process are found: 
$\mu Ti\to eTi$, $Z\to\tau^\mp l^\pm$, $K_L\to e^\mp\mu^\pm$, 
$B^-\to K^{*-}\mu^\mp e^\pm$, $\tau\to e\pi^+K^-$ and 
$B^-\to K^{*-}\tau^\mp e^\pm$. 
The upper bounds of exact BRs are 
evaluated using the improved version of the 
upper bound procedure found in the previous publication \cite{FI96}.
The results are compared with present experimental upper bounds.
For maximal values of model parameters, only six processes have the 
theoretical upper bounds larger than the experimental ones: $\mu\to
e\gamma$, $\mu\to eee$, $\mu Ti\to eTi$, $\tau\to
e\rho^0$, $\tau\to e\pi^+\pi^-$ and $Z\to e\tau$. For these processes
new limits on combinations of $B_{lN}$ matrices are obtained.
The first three have been
studied before \cite{TBBJ95} and they give a new limit on the nondiagonal 
$\mu$-$e$ mixing. The limit is updated here. For the last three, the
ratio of the theoretical and experimental upper bounds are very close to one
and the obtained limit is mass dependent. Therefore, it is not useful.
Two orders od magitude improvement of experimental sensitivities
is needed to obtain mass independent limits on the nondiagonal 
$\tau$-$e$ and $\tau$-$\mu$ mixings from $\tau\to l\gamma$ decays.
Concerning the processes with the $box$-amplitude only, the $K_L\to
e^\mp\mu^\pm$ decay has the best ratio of theoretical to experimental
upper bound. Nevertheless, neutrinoless LFV $B$-meson decays have 
BRs of the order $\sim 10^{-9}$, what makes them 
interesting for finding LFV in future experiments. 
If the structure of the massless part of the $B$ matrix is as
suggested by the Super-Kamiokande experiment, 
one may expect that in the future
the processes containing $\tau$ and $\mu$ leptons in the final (and
initial) state will be most interesting for finding LFV.
Besides BRs for the low-energy neutrinoless LFV decays, 
the constant characteristic for the 
muonium--antimuonium conversion $G_{M\bar{M}}$ is evaluated. The result
obtained is too small to be interesting experimentally.

All above results depend only on the gauge structure of the model 
used and masses of heavy neutrinos. The results do not change if
the massless neutrinos are replaced with the light neutrinos satisfying
present experimental limits. 
A comment on extraction of heavy neutrino mixings from astrophysical
observations is given. 
Following the $V$-model assumption of massless "light" neutrinos,
an  analysis of
oscillations of three 
massless neutrinos in the supernovae is done.
The analysis gives the limits on
mixings in the massless neutrino sector that are in a slight
contradiction with the Super-Kamiokande results.\\

\noindent
{\bf ACKNOWLEDGEMENTS}. Part of this work was  presented on the TAU~'98
Workshop at Santander, Spain. I am indebted to K.K. Gan who promoted my
participation at 
the workshop, to all members of
the organizing committee of the workshop, 
and especially to A. Pich and A. Ruiz who 
invited me to the workshop. Very
pleasant atmosphere, stimulating discussions with K.K. Gan, R. 
Stroynowski and many other colleagues at the workshop stimulated the
present investigation. I am indebted to R. Horvat for many
instructive discussions on neutrino oscillations and supernovae.
I kindly thank S. Fajfer for giving me the best values for 
"CHPT+HQET" parameters needed for neutrinoless LFV decays of heavy
meson. This work is supported by project 1-03-233 "Field theory and
structure of elementary particles".

\appendix
\section{}
The form factors $a_P$ and $b_P$ and $a_V$, $b_V$, $c_V$ and $d_V$ follow
directly from matrix elements of corresponding hadronic currents (\ref{hadr}).
They read
\begin{eqnarray}
\label{HPformf}
a_P&= & -\frac{f_{\cal_H}}{f_P}-2g\,\frac{f_{{\cal H}'^*}}{f_P}\,
\frac{p_2\cdot q}{m^2_{{\cal H}'^*}}\,\frac{(m_1
m^3_{{\cal H}'^*})^{1/2}}{q^2-m^2_{{\cal H}'^*}},
\nonumber\\
b_P&= & 2g\,\frac{f_{{\cal H}'^*}}{f_P}\,\bigg(1+\frac{p_2\cdot
q}{m^2_{{\cal H}'^*}}\bigg)\,\frac{(m_1
m^3_{{\cal H}'^*})^{1/2}}{q^2-m^2_{{\cal H}'^*}}
\end{eqnarray}
and
\begin{eqnarray}
\label{HVformf}
a_V& =& 2^{3/2}\lambda g_V f_{\cal H}'\frac{(m^3_{{\cal H}'^*}m_{\cal
H}^{-1})^{1/2}}{q^2-m^2_{{\cal H}'^*}},
\nonumber\\
b_V& =& -2^{1/2}\beta g_V f_{\cal H}'\frac{(m^3_{{\cal H}'^*}m_{\cal
H}^{-1})^{1/2}}{q^2-m^2_{{\cal H}'^*}},
\nonumber\\
c_V& =& -2^{1/2}\alpha_1 g_V(m_{\cal H}')^{1/2},
\nonumber\\
d_V& =& 2^{1/2}\alpha_2\frac{(m_{\cal H}'^*)^{1/2}}{m_1^2}.
\end{eqnarray}
The phase functions $Z_{Pi}$, $i=1,2,3$ and $Z_{Vi}$, $i=1,\cdots ,8$ 
in the square bracket expressions in (\ref{BRHXll}) read
\begin{eqnarray}
Z_{P1}& =& \int_{s_{13}^{min}}^{s_{13}^{max}}ds_{13}
\, [2p_1\cdot p_3 p_1\cdot p_4-
m_1^2\, p_3\cdot p_4],
\nonumber\\
Z_{P2}& =& \int_{s_{13}^{min}}^{s_{13}^{max}} ds_{13}\, [2(p_1\cdot p_3
p_2\cdot p_4+p_1\cdot p_4p_2\cdot p_3-p_1\cdot p_2 p_3\cdot p_4)],
\nonumber\\
Z_{P3}& =& \int_{s_{13}^{min}}^{s_{13}^{max}} ds_{13}\, 
[2p_2\cdot p_3 p_2\cdot p_4
-m_2^2\, p_3\cdot p_4],
\end{eqnarray}
for ${\cal H}\to Pll'$ decays and
\begin{eqnarray}
Z_{V1}& =& \int_{s_{13}^{min}}^{s_{13}^{max}} ds_{13}\,
[p_1\cdot p_3 p_1\cdot p_2 p_2\cdot p_4
+p_2\cdot p_3 p_1\cdot p_4 p_1\cdot p_2
-m_1^2\, p_2\cdot p_3 p_2\cdot p_4
\nonumber\\&&
-m_2^2\, p_1\cdot p_3 p_1\cdot p_4],
\nonumber\\
Z_{V2}& =& \int_{s_{13}^{min}}^{s_{13}^{max}} ds_{13}\,
\bigg[-q^2 p_3\cdot q p_4\cdot q
+\frac{1}{2}q^2 p_3\cdot p_4
+\frac{1}{m_2^2}(p_2\cdot q)^2\bigg(p_3\cdot q p_4\cdot q 
\nonumber\\&&
-\frac{1}{2}q^2 p_3\cdot p_4\bigg)\bigg],
\nonumber\\
Z_{V3}& =& \int_{s_{13}^{min}}^{s_{13}^{max}} ds_{13}\,
\bigg[p_3\cdot p_4+
\frac{1}{m_2^2}\bigg(p_2\cdot p_3 p_2\cdot p_4
-\frac{1}{2} m_2^2 p_3\cdot p_4\bigg)\bigg],
\nonumber\\
Z_{V4}& =& \int_{s_{13}^{min}}^{s_{13}^{max}} ds_{13}\,
\bigg[-q^2 p_1\cdot p_3 p_1\cdot p_4
+\frac{1}{2} m_1^2 q^2 p_3\cdot p_4
+\frac{1}{m_2^2}(p_2\cdot q)^2\bigg(p_1\cdot p_3 p_1\cdot p_4
\nonumber\\&&
-\frac{1}{2} m_1^2 p_3\cdot p_4\bigg)\bigg],
\nonumber\\
Z_{V5}& =& \int_{s_{13}^{min}}^{s_{13}^{max}} ds_{13}\,
[2p_1\cdot p_3 p_2\cdot p_4
-2p_1\cdot p_4 p_2\cdot p_3],
\nonumber\\
Z_{V6}& =& \int_{s_{13}^{min}}^{s_{13}^{max}} ds_{13}\,
\bigg[-2 p_3\cdot q p_4\cdot q 
+q^2 p_3\cdot p_4
+\frac{1}{m_2^2}p_2\cdot q(p_2\cdot p_3 p_4\cdot q 
\nonumber\\&&
+p_2\cdot p_4 p_3\cdot q 
-p_3\cdot p_4p_2\cdot q)\bigg],
\nonumber\\
Z_{V7}& =& \int_{s_{13}^{min}}^{s_{13}^{max}} ds_{13}\,
\bigg[-q^2 p_3\cdot q p_1\cdot p_4
-q^2 p_4\cdot q p_1\cdot p_3
+q^2 p_1\cdot q p_3\cdot p_4
\nonumber\\&&
+\frac{1}{m_2^2}(p_2\cdot q)^2(p_3\cdot q p_1\cdot p_4+
p_4\cdot q p_1\cdot p_3
-p_1\cdot q p_3\cdot p_4)\bigg],
\nonumber\\
Z_{V8}& =& \int_{s_{13}^{min}}^{s_{13}^{max}} ds_{13}\,
\bigg[-p_3\cdot q p_1\cdot p_4 
-p_4\cdot q p_1\cdot p_3
+p_1\cdot q p_3\cdot p_4
\nonumber\\&&
+\frac{1}{m_2^2}p_2\cdot q(p_2\cdot p_3 p_1\cdot p_4 
+p_2\cdot p_4 p_1\cdot p_3 
-p_1\cdot p_2 p_3\cdot p_4)\bigg],
\end{eqnarray}
for ${\cal H}\to Vll'$ decays. The $p_1$, $p_2$, $p_3$ and $p_4$ 
are 4-momenta of a heavy meson  ($\cal H$), a light meson 
($P$ of $V$), a lepton ($l$) and antilepton ($l'$), respectively.
The corresponding masses are $m_1$, $m_2$, $m_3$ and $m_4$.
The phase functions contain integration over Mandelstam variable 
$s_{13}=(p_1-p_3)^2$. The limits of integration are defined in the 
standard way \cite{PDG98}.

\newpage

Table I. List of neutrinoless LFV processes, the composite loop 
functions and the tree level functions
contributing to them and the approximations (physics) needed
for evaluation of amplitudes. $l$, $P$, $V$, ${\cal H}$ and $B$ 
denote leptons, light pseudoscalar mesons, light vector mesons,
heavy pseudocsalar mesons (containing $c$ or $b$ quark), and 
light baryons, respectively. 
In the first column the list of the neutrinoless
LFV processes is given, with references only 
to the calculations made within
extensions of SM with 
heavy neutrinos. The abbreviations cqf $=$ conserved quark
flavor, ncqf $=$ nonconserved quark flavor and H $=$ Higgs mediated
process, serve to distinguish processes with seemingly similar particle
content. In the second column, the Feynman diagrams 
contributing to any specific process are listed. For instance,
$l$-$q$-$box$ corresponds to the box diagram with one lepton current and
one quark current. In the third column the approximations and physics
used for calculation of amplitudes are listed. Following 
abbreviations are used: HQET $=$ heavy quark effective theory,
CHPT $=$ chiral perturbation theory,
VMD $=$ vector meson dominance, GTR $=$ Goldberger-Treiman relation,
l $=$ lepton physics, q $=$ quark physics.\\

\begin{tabular}{|l|l|l|}
\hline
process  & diagrams & approximations (physics) \\
\hline
$l\to l'\gamma$ \cite{IP_NPB,GGV92} & $\gamma$ & l\\
$\mu\to e$ conversion \cite{TBBJ95,COKFV93,BNT93} &
   $\gamma$, $Z$ and $l$-$q$-$box$ & l, q, nuclear\\
$M\to\bar{M}$ conversion & $l$-$box$ & l, atomic \\
$l^-\to l'^-l_1^-l_2^+$ \cite{IP_NPB,TBBJ95,GGV92} &
   $\gamma$, $Z$ and $l$-$box$ & l \\
$\tau\to lP^0$ (cqf) \cite{IP_NPB,GGV92} &
   $\gamma$, $Z$ and $l$-$q$-$box$ & l, q, PCAC\\
$\tau\to lP^0$ (ncqf) \cite{IP_NPB} &
   $l$-$q$-$box$ & l, q, PCAC\\
$\tau\to lV^0$ (cqf) \cite{IP_NPB} &
   $\gamma$, $Z$ and $l$-$q$-$box$ & l, q, VMD\\
$\tau\to lV^0$ (ncqf) \cite{IP_NPB} &
   $l$-$q$-$box$ & l, q, VMD\\
$Z\to ll'$ \cite{IP_NPB,KPS_Z93} & $\gamma$, $Z$ and $l$-$box$ & l
\\
$H\to ll'$ \cite{P_H92,KPS_H93} & $H$ & l \\
$P^0\to e\mu$ (cqf) \cite{FI96} &
   $\gamma$, $Z$ and $l$-$q$-$box$ & l, q, PCAC\\
$P^0\to e\mu$ (ncqf) \cite{FI96} &
   $l$-$q$-$box$ & l, q, PCAC\\
${\cal H}^0\to ll'$ &
   $l$-$q$-$box$ & l, q, PCAC\\
$\tau^-\to l'^-P_1P_2$ (cqf) \cite{I_PRD96} &
   all except $l$-$box$& l, q, CHPT, PCAC, VMD\\
$\tau^-\to l'^-P_1P_2$ (ncqf) \cite{I_PRD96} &
   $l$-$q$-$box$ and $W^+W^-$ & l, q, CHPT, PCAC, VMD\\
$\tau^-\to l'^-P_1P_2$ (cqf,H) \cite{I_PRD96} &
   $H$ and $W^+W^-$ & l, q, CHPT, PCAC\\
$P_1\to P_2 e\mu$ \cite{FI96} & $l$-$q$-$box$ & l, q, VMD, CHPT\\
${\cal H}\to P l\bar{l}'$  & $l$-$q$-$box$ & l, q, VMD, CHPT, HQET\\
${\cal H}\to V l\bar{l}'$  & $l$-$q$-$box$ & l, q, VMD, CHPT, HQET\\
$B_1\to B_2 e\mu$ \cite{FI96} & $l$-$q$-$box$ & l, q, PCAC, GTR\\
\hline
\end{tabular}\\[0.5cm]

\newpage

Table IIa. Coefficients defining the meson content in axial-vector
quark currents with denoted quark content and normalization given by
Eq. (\ref{cnorm}). 
Two additional coefficients
are different
from zero: $\alpha^{box,ds}_{K^0}=1$ and 
$\alpha^{box,sd}_{\bar{K}^0}=1$.\\

\begin{tabular}{|l|l|l|l|l|}
\hline
$P^0$ & $\alpha^Z_{P^0}$ & $\alpha^{box,uu}_{P^0}$ & 
  $\alpha^{box,dd}_{P^0}$ & $\alpha^{box,ss}_{P^0}$ \\
\hline
$\pi^0$ & $-\sqrt{2}$ 
  & $-\frac{1}{\sqrt{2}}$ 
  & $-\frac{1}{\sqrt{2}}$ 
  & $0$ \\
$\eta$ & $-\frac{\sqrt{2}c_P}{\sqrt{3}}-\frac{s_P}{\sqrt{3}}$ 
  & $-\frac{c_P}{\sqrt{6}}+\frac{s_P}{\sqrt{3}}$ 
  & $ \frac{c_P}{\sqrt{6}}-\frac{s_P}{\sqrt{3}}$ 
  & $-\frac{\sqrt{2}c_P}{\sqrt{3}}-\frac{s_P}{\sqrt{3}}$ \\
$\eta'$ & $\frac{c_P}{\sqrt{3}}-\frac{\sqrt{2}s_P}{\sqrt{3}}$
  & $-\frac{c_P}{\sqrt{3}}-\frac{s_P}{\sqrt{6}}$
  & $ \frac{c_P}{\sqrt{3}}+\frac{s_P}{\sqrt{6}}$
  & $\frac{c_P}{\sqrt{3}}-\frac{\sqrt{2}s_P}{\sqrt{3}}$\\
\hline
\end{tabular}\\[2cm]

Table IIb. Coefficients defining the meson content in vector
quark currents with denoted quark content and normalization given in 
Eq. (\ref{cnorm}). 
Two additional coefficients are different
from zero: $\alpha^{box,ds}_{K^{*0}}=-1$ and
$\alpha^{box,sd}_{\bar{K}^{*0}}=-1$.\\

\begin{tabular}{|l|l|l|l|l|l|l|}
\hline
$V^0$ 
  & $\alpha^Z_{V^0}$ 
  & $\alpha^{box,uu}_{V^0}$ 
  & $\alpha^{box,dd}_{V^0}$
  & $\alpha^{box,ss}_{V^0}$ 
  & $\beta^\gamma_{V^0}$
  & $\gamma^\gamma_{V^0}$\\
\hline
$\rho^0$ & $\sqrt{2}c_{2W}$
  & $\frac{1}{\sqrt{2}}$
  & $\frac{1}{\sqrt{2}}$
  & $0$
  & $2\sqrt{2}s_W^2$ 
  & $-2\sqrt{2}s_W^2$ \\
$\phi$ & $\frac{\sqrt{2}c_Vc_{2W}}{\sqrt{3}}+\frac{s_V}{\sqrt{3}}$
  & $ \frac{c_V}{\sqrt{6}}-\frac{s_V}{\sqrt{3}}$
  & $-\frac{c_V}{\sqrt{6}}+\frac{s_V}{\sqrt{3}}$
  & $\frac{\sqrt{2}c_V}{\sqrt{3}}+\frac{s_V}{\sqrt{3}}$
  & $\frac{2\sqrt{2}c_Vs_W^2}{\sqrt{3}}$
  & $-\frac{2\sqrt{2}c_Vs_W^2}{\sqrt{3}}$\\
$\omega$ & $-\frac{c_V}{\sqrt{3}}+\frac{\sqrt{2}s_Vc_{2W}}{\sqrt{3}}$
  & $ \frac{c_V}{\sqrt{3}}+\frac{s_V}{\sqrt{6}}$
  & $-\frac{c_V}{\sqrt{3}}-\frac{s_V}{\sqrt{6}}$
  & $-\frac{c_V}{\sqrt{3}}+\frac{\sqrt{2}s_V}{\sqrt{3}}$
  & $\frac{2\sqrt{2}s_Vs_W^2}{\sqrt{3}}$
  & $-\frac{2\sqrt{2}s_Vs_W^2}{\sqrt{3}}$\\
\hline
\end{tabular}

\newpage

Table IIIa. The comparison of experimental and theoretical upper bounds
on LFV BRs. Experimental upper bounds for unmarked processes are
taken from Ref. \cite{PDG98}, while those denoted by $^\#$ are from Ref.
\cite{KPJ98}. The newest value 
$B^{UB}(\mu^-\to e^-\gamma)=1.2\times 10^{-11}$ is given 
in Ref. \cite{MDpk99}.\\

\noindent 
\begin{tabular}{|l|l|l|}
\hline
Process  & $B^{UB}_{exp}$ & $B^{UB}_{th}$ \\
\hline
$^\#\mu^-\to e^-\gamma$ & $3.8\times 10^{-11}$ & 
  $8.08\times 10^{-9}x_{\mu e}^2$\\
$\tau^-\to e^-\gamma$ & $2.7\times 10^{-6}$ & 
  $3.38\times 10^{-8}x_{\tau e}^2$\\
$\tau^-\to \mu^-\gamma$ & $3.0\times 10^{-6}$ & 
  $6.68\times 10^{-9}x_{\tau\mu}^2$\\
$\mu^-\to e^-e^+e^-$ & $1.0\times 10^{-12}$ & 
  $6.41\times 10^{-7}y_{\mu e}^2$\\
$\tau^-\to e^-e^+e^-$ & $2.9\times 10^{-6}$ & 
  $2.69\times 10^{-6}y_{\tau e}^2$\\
$\tau^-\to \mu^-\mu^+\mu^-$ & $1.9\times 10^{-6}$ & 
  $4.48\times 10^{-7}y_{\tau\mu}^2$\\
$\tau^-\to e^-\mu^+\mu^-$ & $1.8\times 10^{-6}$ & 
  $1.44\times 10^{-6}y_{\tau e}^2$\\
$\tau^-\to \mu^-e^+e^-$ & $1.7\times 10^{-6}$ & 
  $3.71\times 10^{-7}y_{\tau\mu}^2$\\
$\tau^-\to e^+\mu^-\mu^-$ & $1.5\times 10^{-6}$ & 
  $1.32\times 10^{-9}y_{\tau\mu\mu e}^2$\\
$\tau^-\to \mu^+e^-e^-$ & $1.5\times 10^{-6}$ & 
  $6.67\times 10^{-9}y_{\tau ee\mu}^2$\\
$\tau^-\to e^-\pi^0$ & $3.7\times 10^{-6}$ & 
  $2.77\times 10^{-6}y_{\tau e}^2$\\
$\tau^-\to \mu^-\pi^0$ & $4.0\times 10^{-6}$ & 
  $5.40\times 10^{-7}y_{\tau\mu}^2$\\
$\tau^-\to e^-\eta$ & $8.2\times 10^{-6}$ & 
  $4.01\times 10^{-7}y_{\tau e}^2$\\
$\tau^-\to \mu^-\eta$ & $9.6\times 10^{-6}$ & 
  $7.81\times 10^{-8}y_{\tau\mu}^2$\\
$\tau^-\to e^-\rho^0$ & $2.0\times 10^{-6}$ & 
  $2.70\times 10^{-6}y_{\tau e}^2$\\
$\tau^-\to \mu^-\rho^0$ & $6.3\times 10^{-6}$ & 
  $5.27\times 10^{-7}y_{\tau\mu}^2$\\
$\tau^-\to e^-\phi$ & $6.9\times 10^{-6}$ &
  $2.30\times 10^{-6}y_{\tau e}^2$\\
$\tau^-\to \mu^-\phi$ & $7.0\times 10^{-6}$ & 
  $4.46\times 10^{-7}y_{\tau\mu}^2$\\
$\tau^-\to e^-\pi^+\pi^-$ & $2.2\times 10^{-6}$ & 
  $2.67\times 10^{-6}y_{\tau e}^2$\\
$\tau^-\to \mu^-\pi^+\pi^-$ & $8.2\times 10^{-6}$ & 
  $5.19\times 10^{-7}y_{\tau\mu}^2$\\
$\tau^-\to e^-K^+K^-$ & $6.0\times 10^{-6}$ & 
  $1.07\times 10^{-6}y_{\tau e}^2$\\
$\tau^-\to \mu^-K^+K^-$ & $15\times 10^{-6}$ & 
  $2.07\times 10^{-7}y_{\tau\mu}^2$\\
$\pi^0\to e^-\mu^+$ & $1.72\times 10^{-8}$ & 
  $5.54\times 10^{-15}y_{\mu e}^2$\\
$\eta\to e^-\mu^+$ & $6\times 10^{-6}$ & 
  $1.61\times 10^{-16}y_{\mu e}^2$\\
$Z\to e^\mp\mu^\pm$ & $1.7\times 10^{-6}$ & 
  $3.43\times 10^{-7}y_{\mu e}^2$\\
$^\#Z\to e^\mp\tau^\pm$ & $7.3\times 10^{-6}$ & 
  $8.08\times 10^{-6}y_{\tau e}^2$\\
$^\#Z\to \mu^\mp\tau^\pm$ & $10\times 10^{-6}$ &
  $1.59\times 10^{-6}y_{\tau\mu}^2$\\
\hline
\end{tabular}
\begin{tabular}{|l|l|l|}
\hline
Process  & $B^{UB}_{exp}$ & $B^{UB}_{th}$ \\
\hline
$^\#\mu^-Ti\to e^-Ti$ & $6.1\times 10^{-13}$ & 
  $1.01\times 10^{-5}y_{\mu e}^2$\\
$\tau^-\to e^-K^0$ & $1.3\times 10^{-3}$ & 
  $9.82\times 10^{-16}x_{\tau e}^2$\\
$\tau^-\to \mu^-K^0$ & $1.0\times 10^{-3}$ & 
  $1.93\times 10^{-16}x_{\tau\mu}^2$\\
$\tau^-\to e^-K^{*0}$ & $5.1\times 10^{-6}$ & 
  $2.40\times 10^{-15}x_{\tau e}^2$\\
$\tau^-\to \mu^-K^{*0}$ & $7.5\times 10^{-6}$ & 
  $4.68\times 10^{-16}x_{\tau\mu}^2$\\
$\tau^-\to e^-\bar{K}^{*0}$ & $7.4\times 10^{-6}$ & 
  $2.40\times 10^{-15}x_{\tau e}^2$\\
$\tau^-\to \mu^-\bar{K}^{*0}$ & $7.5\times 10^{-6}$ & 
  $4.68\times 10^{-16}x_{\tau\mu}^2$\\
$\tau^-\to e^-\pi^+K^-$ & $6.4\times 10^{-6}$ & 
  $3.29\times 10^{-15}x_{\tau e}^2$\\
$\tau^-\to \mu^-\pi^+K^-$ & $6.5\times 10^{-6}$ & 
  $6.37\times 10^{-16}x_{\tau\mu}^2$\\
$\tau^-\to e^-\pi^-K^+$ & $3.8\times 10^{-6}$ & 
  $3.29\times 10^{-15}x_{\tau e}^2$\\
$\tau^-\to \mu^-\pi^-K^+$ & $7.4\times 10^{-6}$ & 
  $6.37\times 10^{-16}x_{\tau\mu}^2$\\
$^\#K_L\to e^\mp\mu^\pm$ & $ 2\times 10^{-11}$ & 
  $3.16\times 10^{-14}x_{\mu e}^2$\\
$^\#K_L\to \pi^0e^\mp\mu^\pm$ & $3.2\times 10^{-9}$ & 
  $0$\\
$^\#K^+\to \pi^+e^\mp\mu^\pm$ & $4.0\times 10^{-11}$ & 
  $9.72\times 10^{-16}x_{\mu e}^2$\\
$\bar{B}^0\to e^\mp\mu^\pm$ & $5.9\times 10^{-5}$ &
  $3.07\times 10^{-15}x_{\mu e}^2$\\
$\bar{B}^0\to e^\mp\tau^\pm$ & $5.3\times 10^{-4}$ & 
  $1.61\times 10^{-11}x_{\tau e}^2$\\
$\bar{B}^0\to \mu^\mp\tau^\pm$ & $8.3\times 10^{-4}$ & 
  $3.18\times 10^{-12}x_{\tau\mu}^2$\\
$\bar{B}^0_s\to e^\mp\mu^\pm$ & $4.1\times 10^{-3}$ & 
  $6.11\times 10^{-14}x_{\mu e}^2$\\
$\bar{B}^-\to \pi^-e^\mp\mu^\pm$ & $6.4\times 10^{-3}$ & 
  $8.16\times 10^{-12}x_{\mu e}^2$\\
$\bar{B}^-\to K^-e^\mp\mu^\pm$ & $6.4\times 10^{-3}$ & 
  $1.02\times 10^{-10}x_{\mu e}^2$\\
$\bar{B}^0\to \bar{K}^0e^\mp\mu^\pm$ & $1.8\times 10^{-5}$ & 
  $9.57\times 10^{-11}x_{\mu e}^2$\\
$\tau^-\to e^-\pi^0\pi^0$ & $6.5\times 10^{-6}$ & 
  $4.02\times 10^{-18}y_{\tau e}^2$\\
$\tau^-\to \mu^-\pi^0\pi^0$ & $14\times 10^{-6}$ & 
  $7.91\times 10^{-19}y_{\tau\mu}^2$\\
$\tau^-\to e^-\eta\eta$ & $35\times 10^{-6}$ & 
  $3.16\times 10^{-17}y_{\tau e}^2$\\
$\tau^-\to \mu^-\eta\eta$ & $60\times 10^{-6}$ & 
  $5.94\times 10^{-18}y_{\tau\mu}^2$\\
$\tau^-\to e^-\pi^0\eta$ & $22\times 10^{-6}$ &
  $0$\\
$\tau^-\to \mu^-\pi^0\eta$ & $24\times 10^{-6}$ &
  $0$\\
\hline
\end{tabular}\\[1cm]

\newpage

Table IIIb. Theoretical upper bounds for some interesting BRs, for
which experimental upper bounds have not been found.\\

\begin{tabular}{|l|l|}
\hline
Proces & $B_{th}^{UB}$\\
\hline
$\tau^-\to e^-K^0\bar{K}^0$ & 
 $6.625\times 10^{-7}z_{\tau e}^2$\\
$\tau^-\to \mu^-K^0\bar{K}^0$ & 
 $1.282\times 10^{-7}z_{\tau\mu}^2$\\
$B_s^0\to e^\mp\tau^\pm$ & 
 $3.34\times 10^{-10}z_{\tau e}^2$\\
$B_s^0\to \mu^\mp\tau^\pm$ & 
 $6.62\times 10^{-11}z_{\tau\mu}^2$\\
$B^-\to\pi^-e^\mp\tau^\pm$ & 
 $1.14\times 10^{-10}z_{\tau e}^2$\\
$B^-\to\pi^-\mu^\mp\tau^\pm$ & 
 $2.24\times 10^{-11}z_{\tau\mu}^2$\\
$B^-\to K^-e^\mp\tau^\pm$ & 
 $1.34\times 10^{-9}z_{\tau e}^2$\\
$B^-\to K^-\mu^\mp\tau^\pm$ & 
 $2.63\times 10^{-10}z_{\tau\mu}^2$\\
$\bar{B}^0\to\bar{K}^0e^\mp\tau^\pm$ & 
 $1.26\times 10^{-9}z_{\tau e}^2$\\
$\bar{B}^0\to\bar{K}^0\mu^\mp\tau^\pm$ & 
 $2.48\times 10^{-10}z_{\tau\mu}^2$\\
$\bar{B}_s^0\to\eta e^\mp\mu^\pm$ & 
 $4.24\times 10^{-11}z_{\mu e}^2$\\
$\bar{B}_s^0\to\eta e^\mp\tau^\pm$ & 
 $5.64\times 10^{-10}z_{\tau e}^2$\\
$\bar{B}_s^0\to\eta\mu^\mp\tau^\pm$ & 
 $1.11\times 10^{-10}z_{\tau\mu}^2$\\
$\bar{B}_s^0\to\eta' e^\mp\mu^\pm$ & 
 $1.13\times 10^{-10}z_{\mu e}^2$\\
$\bar{B}_s^0\to\eta' e^\mp\tau^\pm$ & 
 $1.35\times 10^{-9}z_{\tau e}^2$\\
$\bar{B}_s^0\to\eta'\mu^\mp\tau^\pm$ & 
 $2.64\times 10^{-10}z_{\tau\mu}^2$\\
\hline
\end{tabular}
\begin{tabular}{|l|l|}
\hline
Proces & $B_{th}^{UB}$\\
\hline
$B^-\to K^{*-} e^\mp\mu^\pm$ & 
 $1.19\times 10^{-10}z_{\mu e}^2$\\
$B^-\to K^{*-} e^\mp\tau^\pm$ & 
 $1.96\times 10^{-9}z_{\tau e}^2$\\
$B^-\to K^{*-}\mu^\mp\tau^\pm$ & 
 $3.85\times 10^{-10}z_{\tau\mu}^2$\\
$\bar{B}^0\to K^{*0} e^\mp\mu^\pm$ &
 $1.12\times 10^{-10}z_{\mu e}^2$\\
$\bar{B}^0\to K^{*0} e^\mp\tau^\pm$ &
 $1.82\times 10^{-9}z_{\tau e}^2$\\
$\bar{B}^0\to K^{*0}\mu^\mp\tau^\pm$ &
 $3.60\times 10^{-10}z_{\tau\mu}^2$\\
$\bar{B}_s^0\phi e^\mp\mu^\pm$ & 
 $1.01\times 10^{-10}z_{\mu e}^2$\\
$\bar{B}_s^0\phi e^\mp\tau^\pm$ & 
 $1.56\times 10^{-9}z_{\tau e}^2$\\
$\bar{B}_s^0\phi\mu^\mp\tau^\pm$ & 
 $3.06\times 10^{-10}z_{\tau\mu}^2$\\
$\Sigma^+\to pe^\mp\mu^\pm$ &
 $4.09\times 10^{-18}z_{\mu e}^2$\\
$\Lambda\to ne^\mp\mu^\pm$ &
 $2.74\times 10^{-18}z_{\mu e}^2$\\
$\Xi^0\to\Lambda e^\mp\mu^\pm$ &
 $3.18\times 10^{-18}z_{\mu e}^2$\\
$\Xi^0\to\Sigma^0e^\mp\mu^\pm$ &
 $1.29\times 10^{-20}z_{\mu e}^2$\\
$\Xi^-\to\Sigma^-e^\mp\mu^\pm$ &
 $2.02\times 10^{-20}z_{\mu e}^2$\\
$\Sigma^0\to ne^\mp\mu^\pm$ &
 $1.99\times 10^{-27}z_{\mu e}^2$\\
$$ &
 $$\\
\hline
\end{tabular}


\begin{figure}
FIG. 1. The BRs and the upper bounds of the BRs
for four leading processes of the groups of processes given 
in equations (\ref{FZmue}), (\ref{FZtaul}), (\ref{Fboxmuesd}) 
and (\ref{Fboxtaulbs}).
Each of these four processes is 
shown in one of four panels. The BRs are evaluated for
degenerate heavy neutrino masses, $m_N$. The model parameters are
adjusted so that the maximal 
BR values are smaller than the present experimental upper bounds. 
It is assumed that the "angle" parameter $x_{\mu\tau}\approx 1$, 
in accord with the Super-Kamiokande measurements.
The full line represents the upper
bound calculation keeping the parameters $x_{ll'}$ equal to 
one, and adjusting the $s_L^{\nu_l}$ parameters: 
$(s_L^{\nu_e})^2=4.29\times 10^{-10}$, 
$(s_L^{\nu_\mu})^2=1.4\times 10^{-3}$, and
$(s_L^{\nu_\tau})^2=3.3\times 10^{-2}$. The heavy long-dashed
line represents the upper bound calculation keeping the $s_L^{\nu_l}$
equal to the present experimental upper bound values (\ref{sL}),
and adjusting the $x_{ll'}$ parameters: 
$x_{\mu e}=x_{\tau e}=2.459\times 10^{-4}$,
$x_{\tau\mu}=1$. The dotted line represents the calculation with the 
"realistic" $B_{lN}$ matrix elements, and with the same parameters as
for the full line calculation.
\end{figure}


\begin{figure}
FIG. 2. The BRs and upper bounds for BRs for the same processes 
as in Fig 1., but now evaluated as a function of the ratio 
of two heavy neutrino masses, $m_{N_2}/m_{N_1}$. 
For all curves the first and third masses are taken to be degenerate,
$m_{N_1}=m_{N_3}=4000\ GeV$, 
while the second mass assumes values within the interval
$1\leq m_{N_2}/m_{N_1}\leq 10^5$. The types of lines
represent the same sets of parameters 
as in Fig. 1. In the first panel, representing the 
$\mu\to e$ conversion on $Ti$, additional curve is added, to show that 
one can always achieve theoretical values smaller than the present
experimental bounds. The calculation for that curve was made within the
upper bound procedure, for $x_{ll'}=1$, 
$(s_L^{\nu_e})^2=(s_L^{\nu_\mu})^2=0.5\times 10^{-10}$ and
$(s_L^{\nu_\tau})^2=0.033$. 
\end{figure}


\begin{thebibliography}{99}
\bibitem{tHooft76} G. t'Hooft, Phys. Rev. Lett. {\bf 37}, 8 (1976).
\bibitem{SupKam98} Super-Kamiokande Collaboration (Y. Fukuda et al.), 
Phys. Rev. Lett. 81 1562, (1998);
Super-Kamiokande Collaboration (T. Kajita), hep-ex/9810001.
\bibitem{DL95} P. Depommier and C. Leroy, Rep. Prog. Phys. {\bf 58},
61 (1995).
\bibitem{KLV94} T.S. Kosmas, G.K. Leontaris and J.D. Vergados,
Prog. Part. Nucl. Phys. {\bf 33}, 397 (1994).
\bibitem{LFV_SMext} R.N. Mohapatra, Prog. Part. 
Nucl. Phys. {\bf 26}, 1 (1991);
J.W.F. Valle, Prog. Part. Nucl. Phys. {\bf 26}, 91 (1991);
J.L. Hewett, T. Takeuchi and S. Thomas, hep-ph/9603391,
In "{\it Barklow, T.L. (ed.) et al.: Electroweak symmetry breaking and new
physics at the TeV scale}" 548-649.
\bibitem{V89} J.W.F. Valle, Nucl. Phys. Proc. Suppl. {\bf 11}, 118
(1989).
\bibitem{AP_ZPC92} A. Pilaftsis, Z. Phys. C {\bf 55}, 275 (1992).
\bibitem{MV86} R.N. Mohapatra and J.W.F. Valle, Phys. Rev. D {\bf 34}, 1642 
(1986);
J.W.F. Valle, UAB-FT-153, {\it 
International Symposium on Nuclear Beta Decays and
Neutrino}, Osaka, June 1986, published in {\it Osaka Beta Symposium} 1986:0499
(QCD175:I65:1986).
\bibitem{KP96} B.A. Kniehl and A. Pilaftsis, Nucl. Phys. {\bf B474},
286 (1996).
\bibitem{FI96} S. Fajfer, A. Ilakovac, Phys. Rev. D {\bf 57}, 4219
(1996).
\bibitem{IP_NPB} A. Ilakovac and A. Pilaftsis, Nucl. Phys. {\bf B437}, 491
(1995).
\bibitem{TBBJ95} D. Tommasini, G. Barenboim, J. Bernab\'eu and C.
Jarlskog, Nucl. Phys. {\bf B444}, 451 (1995).
\bibitem{NRT94} E. Nardi, E. Roulet and D. Tommasini, 
Phys. Lett. B {\bf 327}, 319 (1994).
\bibitem{NQRV96} H. Nunokawa, Y.-Z. Qian, A. Rossi and J.W.F. Valle,
Phys. Rev. D {\bf 54}, 4356 (1996).
\bibitem{SY91} M. Sher and Y. Yuan, Phys. Rev. D {\bf 44}, 1461 (1991).
\bibitem{BW86} W. Buchm\"uller and D. Wyler, Phys. Lett. B {\bf 177}, 377
(1986).
\bibitem{JOE90} J.O. Eeg, Z. Phys. C {\bf 46}, 665 (1990).
\bibitem{GGV92} M.C. Gonzalez-Garcia and J.W.F. Valle, Mod. Phys. Lett.
A {\bf 7} 477 (1992);
Mod. Phys. Lett. A {\bf 9}, 2569 (1994).
\bibitem{BSVMV87} J. Bernab\'eu, A. Santamaria, J. Vidal, A. Menez and
J.W.F. Valle, Phys. Lett. B {\bf 187}, 303 (1987).
\bibitem{GGV89} M.C. Gonzalez-Garcia and J.W.F. Valle, Phys. Lett. B
{\bf 216}, 360 (1989).
\bibitem{WW83} D. Wyler and L. Wolfenstein, Nucl. Phys. {\bf B218}, 205 (1983).
\bibitem{Wi86} E. Witten, Nucl. Phys. {\bf B268}, 79 (1986).
\bibitem{RV90} N. Rius and J.W.F. Valle, Phys. Lett. B {\bf 246}, 249 (1990).
\bibitem{BRV89} G.C. Branco, M.N. Rebelo and J.W.F. Valle,
Phys. Lett. B {\bf 225}, 385 (1989).
\bibitem{DSGV90} M. Dittmar, A. Santamaria, M.C. Gonzalez-Garcia and
J.W.F. Valle, Nucl. Phys. {\bf B332}, 1 (1990).
\bibitem{LL88_2} P. Langacker and D. London, Phys. Rev. D {\bf 38}, 907 (1988).
\bibitem{LL88} P. Langacker and D. London, Phys. Rev. D {\bf 38}, 886 (1988).
\bibitem{NRT92} E. Nardi, E. Roulet and D. Tommasini, Nucl. Phys. {\bf B386},
239 (1992).
\bibitem{BDGR91} G. Bhattacharyya, A. Datta, S.N. Ganguli and
Raychaudhuri, Mod. Phys. Lett. A {\bf 6}, 2921
(1991).
\bibitem{BGKLM94} C.P. Burgess, S. Godfrey, H. K\"onig, D. London and I.
Maksymyk, Phys. Rev. D {\bf 49}, 6115 (1994).
\bibitem{BKM95} G. Bhattacharyya, P. Kalyniak and I. Melo, Phys. Rev. D
{\bf 51}, 3569 (1995).
\bibitem{KM97} P. Kalyniak and I. Melo, Phys. Rev. D {\bf 55}, 1453 (1997).
\bibitem{LFVtheo} 
P. Minikowski, Phys. Lett. B {\bf 67}, 421 (1977);
E. Ma and A. Pramudita, Phys. Rev. D {\bf 22}, 214 (1980);
B. McWilliams and L.-F. Li, Nucl. Phys. {\bf B179}, 62 (1981);
S. Dimopoulos and J. Ellis, Nucl. Phys. {\bf B182}, 505 (1981);
S.T. Petcov, Phys. Lett. B {\bf 115}, 401 (1982);
J.C. Pati, Phys. Rev. D {\bf 30}, 1144 (1984);
O.W. Greenberg, R.N. Mohapatra and S. Nussinov, Phys. Lett. B {\bf 148},
465 (1984);
P. Herceg and T. Oka, Phys. Rev. D {\bf 29}, 475 (1984);
W.-S. Hou and A. Soni, Phys. Rev. Lett. {\bf 54}, 2083 (1985);
J.C. Pati and H. Stremintzer, Phys. Lett. B {\bf 172}, 441 (1986);
V. Barger, G.F. Giudice and T. Han, Phys. Rev. {\bf 40}, 2987 (1989);
S.F. King, Nucl. Phys. {\bf B320}, 487 (1989);
R. Arnowith and P. Nath, Phys. Rev. Lett. {\bf 66}, 2708 (1991);
S. Kelley, J.L. Lopez, D.V. Nanopoulos and H. Pois, Nucl. Phys. 
{\bf B358}, 27 (1991);
J.C. Rom\~ ao, N. Rius and J.W.F. Valle, Nucl. Phys. {\bf B363}, 369
(1991);
R.N. Mohapatra, Phys. Rev. D {\bf 46}, 1992 (1990);
J. Wu, S. Urano and R. Arnowitt, Phys. Rev. D {\bf 47}, 4006 (1993);
L.N. Chang, D. Ng and J.N. Ng, Phys. Rev. D {\bf 50}, 4589 (1994);
R.N. Mohapatra, S. Nussinov and X. Zhang, Phys. Rev. D {\bf 49}, 2410
(1994);
T. Kosmas, G. Leontaris and J. Vergados, Prog. Part. Nucl. Phys. 
{\bf 33}, 397 (1994);
R. Barbieri, L. Hall and A. Strumia, Nucl. Phys. {\bf B455}, 219 (1995);
M. Frank and H. Hamidian, Phys. Rev. D {\bf 54}, 6790 (1996);
N. Arkani-Hamed, H.-C. Cheng and L.J. Hall, Phys. Rev. D {\bf 53}, 413
(1996);
G. Barneboim and M. Raidal, Nucl. Phys. {\bf B484}, 63 (1997);
Z. Gagyi-Palffi, A. Pilaftsis and K. Schilcher, Nucl. Phys. {\bf B513},
517 (1998);
M. Raidal and A. Santamaria, Phys. Lett. B {\bf 421}, 250 (1998);
R. Kitano and K. Yamamoto, hep-ph/9905459.
\bibitem{tauexp} CLEO Collaboration (T. Bowcock et al.), 
Phys. Rev. D {\bf 41}, 805 (1990); 
ARGUS Collaboration (H. Albrecht et al.),
Z. Phys. C {\bf 55}, 179 (1992); 
CLEO Collaboration (J. Bartelt et al.), 
Phys. Rev. Lett. {\bf 73}, 1890 (1994); 
K.K. Gan, Nucl. Phys. Proc. Suppl. {\bf 55 C}, (1997) 213;
CLEO Collaboration (K.W. Edwards et al.),
Phys. Rev. D {\bf 55}, R3919 (1997); 
CLEO Collaboration (G. Bonvicini et al.),
Phys. Rev. Lett. {\bf 79}, 1221 (1997); 
CLEO Collaboration (D.W. Bliss et al.),
Phys. Rev. D {\bf 57}, 5903 (1998).
\bibitem{muexp} For reviews see A. van der Schaff, Prog. Part. Nucl.
Phys. {\bf 31}, 1 (1993); 
A. Czarnecki, In "{\it Balholm 1997, Beyond the standard model
5}", 252, hep-ph/9710425.
\bibitem{PDG98} Particle Data Group, 
(C. Caso et al.), Eur. Phys. J. C {\bf 3}, 1 (1998).
\bibitem{KPJ98} K.P. Jungmann,
hep-ex/9806003.
\bibitem{V87} J.W.F. Valle, Phys. Lett. B {\bf 199}, 432 (1987).
\bibitem{Si95} G. Sigl, Phys. Rev. D {\bf 51}, 4035 (1995);
G. Raffelt and G. Sigl, Astropart. Phys. {\bf 1}, 164 (1993).
\bibitem{LZapp} L.D. Landau Phys. Z. Sow. {\bf 2}, 46 (1932);
C. Zener, Proc. R. Soc. London {\bf 137A}, 696 (1932).
\bibitem{KP89} T.K. Kuo and J. Pantaleone, Rev. Mod. Phys. {\bf 61}, 937 (1989).
\bibitem{SSB94} A.Yu. Smirnov, D.N. Spergel and J.N. Bahcall, Phys. Rev.
D {\bf 49}, 1389 (1994).
\bibitem{KP88} T.K. Kuo and J. Pantaleone, Phys. Rev. D {\bf 37}, 298 (1988).
\bibitem{AC} T. Appelquist and J. Carazzone, Phys. Rev. D {\bf 11},
2856 (1975).
\bibitem{SS} G. Senjanovi\'c and A. \v Sokorac, Nucl. Phys. {\bf B164}, 
305 (1980).
\bibitem{XYP98} X.-Y. Pham, Eur. Phys. J. C {\bf 8}, 513,(1999).
\bibitem{GGetal98} M.C. Gonzalez-Garcia et al., Phys. Rev. Lett. {\bf 82},
3202 (1999). 
\bibitem{FLMS99} G.L. Fogli, E. Lisi, A. Marrone and G. Scioscia,
Phys. Rev. D {\bf 60}, 053006 (1999). 
\bibitem{FGGV99} N. Fornengo, M.C. Gonzalez-Garcia and 
J.W.F. Valle, hep-ph/9906539.
\bibitem{IKP95} A. Ilakovac, B.A. Kniehl and A. Pilaftsis, Phys. Rev. D {\bf
52}, 3993 (1995).
\bibitem{Cas97} R. Casalbuoni, A. Deandrea, N.Di Bartolomeo, G. Gatto, 
F.Feruglio and G. Nardulli, Phys. Rept. {\bf 281}, 145 (1997).
\bibitem{BFO95} B. Bajc, S. Fajfer and R.J. Oakes, Phys. Rev. D {\bf 51},
2230 (1995).
\bibitem{HQ_CH} M.B. Wise, Phys. Rev. D {\bf 45}, R2188 (1992);
G. Burdman and J.F. Donoghue; Phys. Lett. B {\bf 280}, 287 (1992).
\bibitem{BFO96a} B. Bajc, S. Fajfer and R.J. Oakes, Phys. Rev. D {\bf 53},
4957 (1996);
\bibitem{BFO96b} B. Bajc, S. Fajfer and R.J. Oakes, Phys. Rev. D {\bf 54},
5883 (1996);
\bibitem{BFOP97} B. Bajc, S. Fajfer and R.J. Oakes, S. Prelov\v sek,
Phys. Rev. D {\bf 56}, 7207 (1997).
\bibitem{BKY88} M. Bando, T. Kugo, S. Uehara, K. Yamawaki and T.
Yanagida, Phys. Rev. Lett. {\bf 54}, 1215 (1985);
M. Bando, T. Kugo, K. Yamawaki,
Nucl. Phys. {\bf B259}, 493 (1985);
M. Bando, T. Kugo, K. Yamawaki,
Phys. Rept. {\bf 164}, 217 (1988).
\bibitem{Cas92} R. Casalbuoni, A. Deandrea, N.Di Bartolomeo, G. Gatto,
F.Feruglio and G. Nardulli, Phys. Lett. B {\bf 294}, 106 (1992);
Phys. Lett. B {\bf 292}, 371 (1992).
\bibitem{Dgamma} P. Cho and H. Georgi, Phys. Lett. B {\bf 296}, 408 (1992);
J.F. Amundson et al., Phys. Lett.
B {\bf 296}, 415 (1992).
\bibitem{FW61} G. Feinberg and S. Weinberg, Phys. Rev. {\bf 123}, 1439 (1961);
G. Feinberg and S. Weinberg, Phys. Rev. Lett. {\bf 6}, 381 (1961).
\bibitem{Po57} B. Pontecorvo, Zh. Eksp. Teor. Fiz. {\bf 33}, 549 (1957).
\bibitem{LW99}
L. Willmann et al., Phys. Rev. Lett. 82, 49 (1999).
\bibitem{M91G94} B.E. Matthias et al. Phys. Rev. Lett. {\bf 66}, 2716 (1991);
V.A. Gordeev et al., JEPT Lett. {\bf 59}, 589 (1994).
\bibitem{MLS89} M.L. Swartz, Phys. Rev. D {\bf 40}, 1521 (1989).
\bibitem{MMmodels} V. Barger, H. Baer, W.Y. Keung and R.J.N. Phillips,
Phys. Rev. D {\bf 26}, 218 (1982);
A. Halprin, Phys. Rev. Lett. {\bf 48}, 1313 (1982);
G.K. Leontaris, K. Tamvakis and J.D. Vergados, Phys. Lett. B {\bf 162},
153 (1985);
Phys. Lett. B {\bf 171}, 412 (1986);
D. Chang and W.-Y. Keung, Phys. Rev. Lett. {\bf 62}, 2583 (1989);
P. Herceg and R.N. Mohapatra, Phys. Rev. Lett. {\bf 69}, 2475 (1992);
R.N. Mohapatra, Z. Phys. C {\bf 56}, S117 (1992);
A. Halprin and A. Masiero, Phys. Rev. D {\bf 48}, R2987 (1993);
H. Fujii, Y. Mimura, K. Sasaki and T. Sasaki, 
Phys. Rev. D {\bf 49}, 559 (1994);
G.-G. Wong and W.-S. Hou Phys. Rev. D {\bf 50}, R2962 (1994);
Phys. Rev. D {\bf 53}, 1537 (1996).
\bibitem{I_PRD96} A. Ilakovac, Phys. Rev. D {\bf 54}, 5653 (1996).
\bibitem{KPS_Z93} J.G. K\"orner, A. Pilaftsis and K. Schilcher, Phys.
Lett. B {\bf 300}, 381 (1993).
\bibitem{P_H92} A. Pilaftsis, Phys. Lett. B {\bf 285}, 68 (1992).
\bibitem{KPS_H93} J.G. K\"orner, A. Pilaftsis and K. Schilcher, Phys.
Rev. D {\bf 47}, 1080 (1993).
\bibitem{COKFV93} H.C. Chiang, E. Oset, T.S. Kosmas, A. Faessler and
J.D. Vergados, Nucl. Phys. {\bf A559}, 526 (1993).
\bibitem{BNT93} J. Bernabeu, E. Nardi and D. Tommasini, Nucl. Phys.
{\bf B409}, 69 (1993). 
\bibitem{FW62} K.W. Ford and J.G. Wills, Nucl. Phys. {\bf 35 }, 295 (1962).
\bibitem{JCS59} J.C. Sens, Phys. Rev. {\bf 113}, 679 (1959).
\bibitem{SMC70} I. Sick and J.S. McCarthy, Nucl. Phys. {\bf A150}, 631 (1970).
\bibitem{DFMRL74} B. Deher, J. Friedrich, K. Merle, H. Rothhaas and G.
L\"uhrs, Nucl. Phys. {\bf A235}, 219 (1974).
\bibitem{SMR87} T. Suzuki, D.F. Measday and J.P. Roalsvig, Phys. Rev. C
2212 (1987);
see also D.A. Bryman et al., Phys. Rev. Lett. {\bf 55}, 465 (1985).
\bibitem{MR97} M. Raidal, hep-ph/9712259.
\bibitem{FWW80ACS81} R. Fischer, J. Wess and F. Wagner, Z. Phys. C {\bf 3},
313 (1980);
G.J. Aubrecht, N. Chahrouri and K. Slanec, Phys. Rev. D {\bf 24},
1318 (1981).
\bibitem{Mir9396} D. Decker, E. Mirkes, R. Sauer and Z. Was, Z. Phys. C
{\bf 58}, 445 (1993);
M. Finkemeier and E. Mirkes, {\it ibid.} {\bf 69}, 243
(1996).
\bibitem{KPP8488} G. Kramer, W.F. Palmer and S.S. Pinsky, Phys. Rev. D
{\bf 30}, 89 (1984);
G. Kramer, W.F. Palmer Z. Phys. C {\bf 25}, 195 (1984);
{\bf 39}, 423 (1988).
\bibitem{BBGCFG} W.A. Bardeen, A.J. Buras and J.-M. G\'erard, Nucl.
Phys. {\bf B293}, 787 (1987);
W.A. Bardeen, A.J. Buras and J.-M. G\'erard,
Phys. Lett. B {\bf 180}, 133 (1986);
R.S. Chivukula, J.M. Flynn and H. Georgi,
{\it ibid.} {\bf 171}, 453 (1986).
\bibitem{PDG94} Particle Data Group, 
(L. Montanet et al.), Phys. Rev. D {\bf 50}, 1173 (1994).
\bibitem{thetaP} CELLO Colaboration (H.-J. Behrend et al.), Z. Phys. C
{\bf 49}, 401 (1991);
TPC Collaboration (H. Aihara et al.), Phys, Rev. Lett.
{\bf 64}, 172 (1990).
\bibitem{FPS98} S. Fajfer, S. Prelov\v sek and P. Singer, Phys. Rev. D
{\bf 58}, 094038 (1998).
\bibitem{F99} S. Fajfer, private communication.
\bibitem{MDpk99} M.L. Brooks et al., hep-ex/9905013, submitted to
Phys. Rev. Lett.

\end{thebibliography}
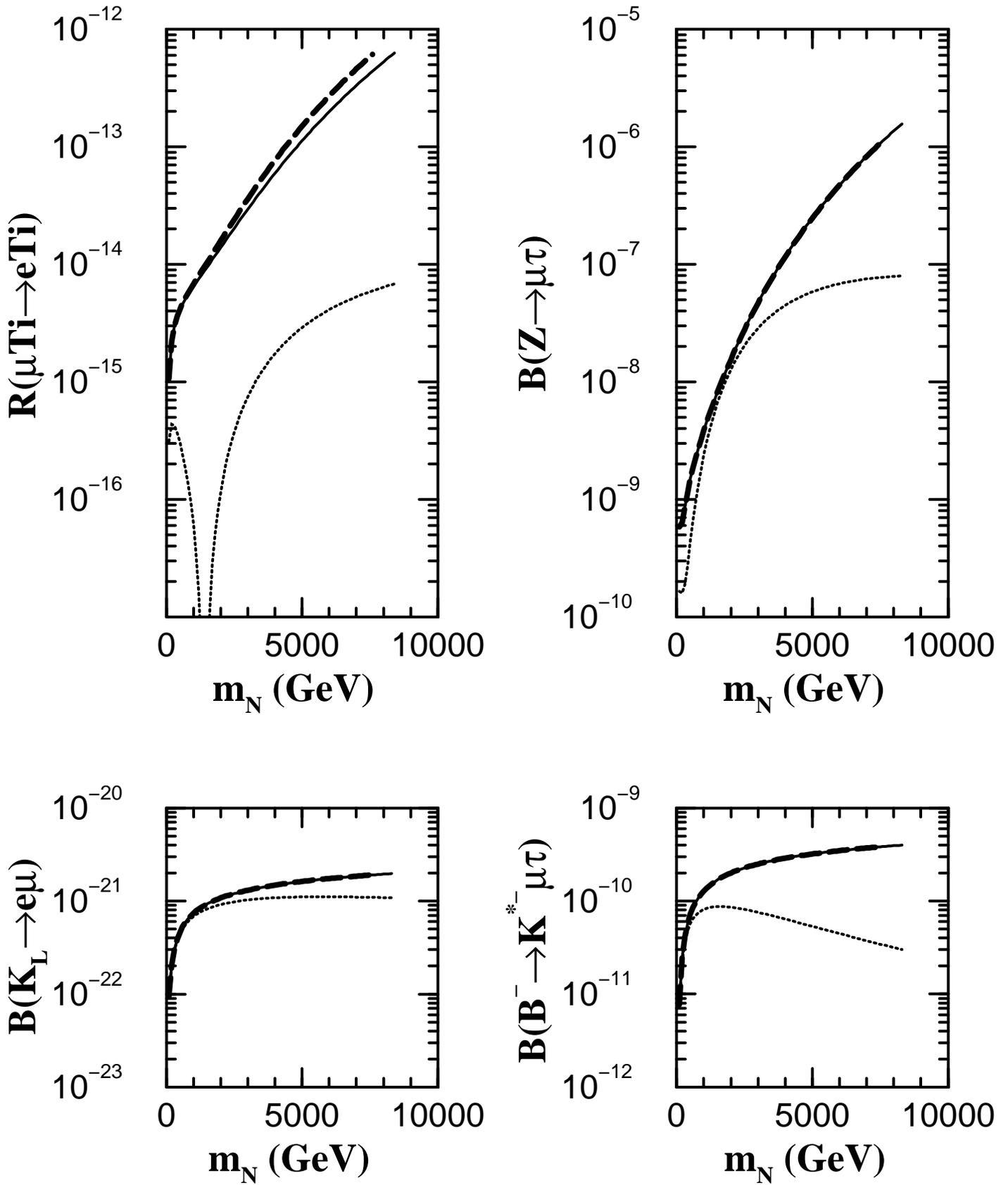
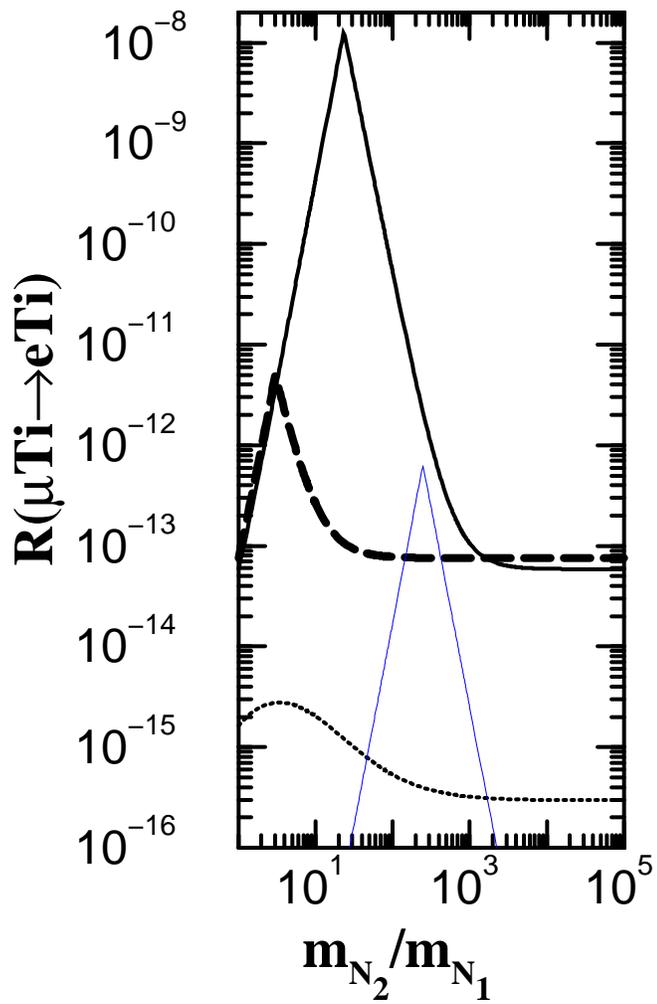
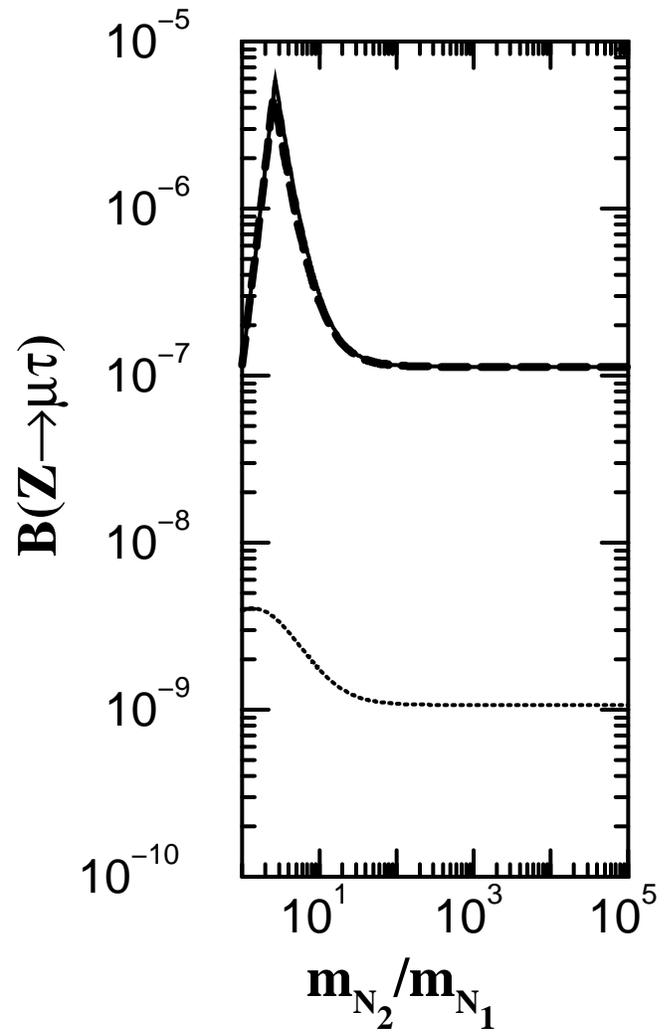
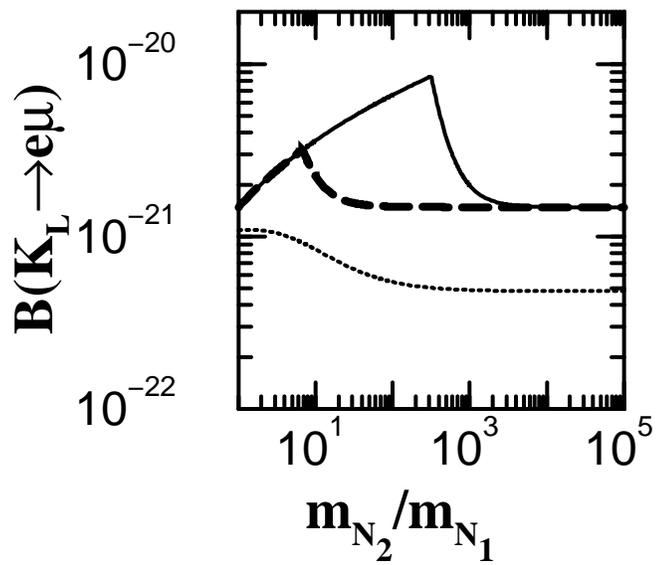
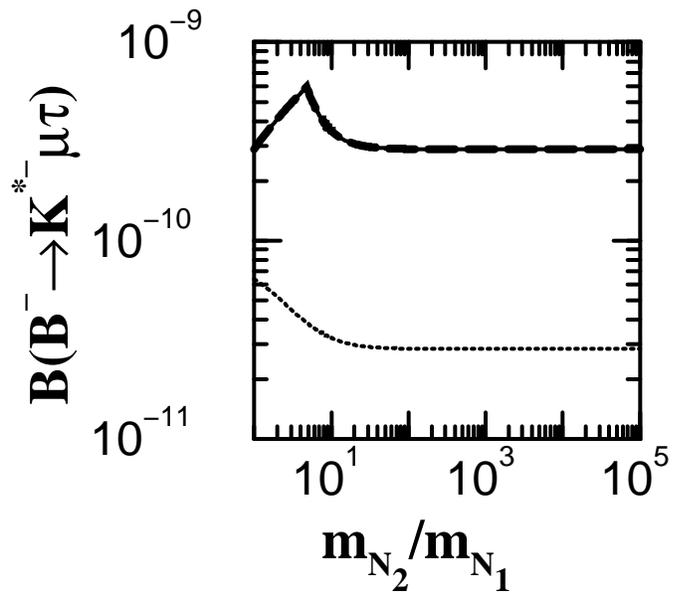
\end{document}